\documentclass[12pt]{article}

\usepackage[a4paper,left=30mm,right=20mm,top=20mm,bottom=20mm]{geometry}
\usepackage{graphics}
\usepackage{amsmath}
\usepackage{amssymb}
\usepackage{epsfig}
\usepackage{cite}
\usepackage{xcolor}
\usepackage[unicode]{hyperref}
\title{TMD parton showers for associated $\gamma$~$+$~jet production in electron-proton collisions at high energies}

\author{A.V.~Lipatov$^{1,2}$, M.A.~Malyshev$^{1}$}

\begin{document}

\maketitle

\begin{center}

{\it $^{1}$Skobeltsyn Institute of Nuclear Physics, Lomonosov Moscow State University, 119991, Moscow, Russia}\\
{\it $^{2}$Joint Institute for Nuclear Research, 141980, Dubna, Moscow region, Russia}\\

\end{center}

\vspace{0.5cm}

\begin{center}

{\bf Abstract }

\end{center}

An earlier developed $k_T$-factorization framework to calculate the 
associated prompt photon and hadronic jets production cross sections at high energies is extended now to the electron-proton deep inelastic scattering.
The proposed method is based on joint usage of the \textsc{pegasus} and \textsc{cascade} Monte-Carlo 
event generators, which deal with the transverse momentum dependent (TMD)
gluon and quark densities in a proton. First of them, \textsc{pegasus}, is applied to produce off-shell 
photon-gluon fusion events. 
Then, to properly simulate the kinematics of the produced jets, the TMD parton 
shower algorithm implemented into the \textsc{cascade} is used. 
We demonstrate basic features of the approach considering latest H1 and ZEUS data on 
prompt photon plus jet production at low $Q^2$ collected at HERA.
We achieve a good description of the measurements and point out that a correct simulation 
of parton showers is essential for studying non-inclusive processes at high energies 
within the TMD-based approaches. 

\indent

\vspace{1cm}

\noindent
{\it Keywords:} deep inelastic scattering, high energy factorization, CCFM evolution, TMD gluon density, prompt photon production

\newpage

\section{Introduction} \indent

Inclusive prompt photon\footnote{Photons are considered as prompt if they are produced in the hard interaction 
subprocess rather than in hadron decays or via fragmentation.} production at high energies is an important 
subject of investigation at modern colliders, such as LHC. 
Being quite clean in sense of final state effects, 
these processes can serve as an important source of information about the proton 
structure (expressed in terms of parton density functions)
and represent an important background to many processes 
involving photons in the final state, including Higgs boson production. 
Also, a rather clear and robust way to study parton content of a 
proton can be provided by the prompt photon photoproduction in
deep inelastic $ep$ scattering (DIS), where
the exchanged photon is of small
virtuality $Q^2 \lesssim 1$~GeV$^2$\cite{PromptPhoton-H1, PromptPhoton-ZEUS1, PromptPhoton-ZEUS2}. In fact, the 
corresponding cross section at leading order (LO) of 
perturbative QCD is mainly 
governed by the direct $\gamma + q \to \gamma + q$ and several resolved photon subprocesses, where
the photon emitted by the electron fluctuates into a hadronic state. A
gluon and/or a quark of this hadronic fluctuation takes part then in hard interactions\footnote{Beyound the LO, the 
difference between these two mechanisms disappears.}.
Including into the consideration accompanying final state jet(s) gives an 
opportunity to study jet observables, different photon-jet correlations, thus 
allowing to test the production dynamics in many additional aspects.
Investigation of prompt photon production will be 
an important part of the physical program at future 
electron-proton or electron-ion colliders, such as LHeC\cite{LHeC}, FCC-eh\cite{FCC}, EiC\cite{EiC} and EiCC\cite{EiCC}.
Thus, relevant theoretical studies and developing corresponding Monte-Carlo event 
generators are rather important tasks at present. 

Usually, a theoretical description of prompt photon photoproduction relies on  
conventional (collinear) QCD factorization, which is based on well-known Dokshitzer-Gribov-Lipatov-Altarelli-Parisi (DGLAP) quark and gluon
evolution equations\cite{DGLAP}. 
Within this formalism, next-to-leading order (NLO) calculations are available\cite{FGH1,FGH2, ZK}.  
The prompt photon production events can be produced also with some Monte-Carlo 
generators, such as \textsc{pythia}\cite{pythia} or \textsc{herwig}\cite{herwig}. 
Alternative calculations can be done in the framework of high-energy factorization\cite{HighEnergyFactorization} 
or $k_T$-factorization approach\cite{kt-factorization}
based on Balitsky-Fadin-Kuraev-Lipatov (BFKL)\cite{BFKL} or 
Catani-Ciafaloni-Fiorani-Marchesini (CCFM)\cite{CCFM} gluon 
evolution equations.
The $k_T$-factorization QCD approach has certain technical advantages in the ease of 
including higher-order pQCD radiative corrections corresponding to
real initial-state gluon emissions (namely, terms, proportional to $\alpha_s^n \ln^n s/\Lambda_{\rm QCD}^2 \sim \alpha_s^n \ln^n 1/x$, important at 
high energies $s$, or small $x$) in the form of transverse momentum dependent (TMD,
or unintegrated) gluon density (see review\cite{TMD-review} for more information).

Studies of prompt photon production within the $k_T$-factorization approach have their own 
long history (see, for example,\cite{LZ-photo1, LZ-photo2, LMZ-photo, Szczurek-photon, BLZ, KNS-photo, LZ-PP, LM-PP, LMJ-PP, Motyka-photon} and references therein).
Moreover, such processes have been also investigated\cite{CGC-photo1, CGC-photo2, CGC-photo3, CGC-photo4, CGC-photo5, CGC-photo6, CGC-photo7,CGC-photo8,CGC-photo9} in the 
framework of Color Glass Condensate (CGC) formalism\cite{CGC-1, CGC-2, CGC-3}.
In particular, a good description of experimental data for inclusive 
production events was obtained for both electron-proton\cite{LZ-photo1, LZ-photo2, LMZ-photo} and proton-proton
collisions\cite{LM-PP, LMJ-PP, Motyka-photon}. 
A simple model\cite{PartonShowerModel} to reconstruct the jet kinematics
in associated $\gamma + {\rm jet}$ production was applied\cite{LZ-PP}
and necessity for correct reconstruction of this kinematics was pointed out. 
The latter is a special task for any calculations 
where the non-collinear QCD evolution of the initial parton (gluon) cascade is used.
In the case of CCFM evolution, a method based on numerical simulation 
of TMD parton showers was proposed\cite{JetReconstructionMethod} and successfully
applied to associated $\gamma + {\rm jet}$ production at LHC\cite{LMJ-PP}.
In fact, earlier description of the experimental data 
was significantly improved.
Here we 
extend the consideration to electron-proton collisions
and analyse latest HERA data\cite{PromptPhoton-H1, PromptPhoton-ZEUS1, PromptPhoton-ZEUS2} on prompt photon and 
associated jet photoproduction taken by the H1 and ZEUS Collaborations at $\sqrt s = 319$~GeV.

The paper is organized as follows. In Section 2 we briefly describe our approach.
In Section~3 we present our numerical results and compare them with available HERA data.
Section~4 contains our conclusions.

\section{Theoretical framework} \indent

Let us shortly describe the main calculation steps.
In the photoproduction regime 
the colliding electron emits a quasi-real ($Q^2 \sim 0$) photon, so one 
deals then with photon-proton interaction:
\begin{gather}
  d\sigma(ep \to e'\gamma X) = \int f_{\gamma/e}(y)d\sigma(\gamma p\to\gamma X)dy,
  \label{eq:DIS}
\end{gather}
\noindent
where $y$ is the fraction of initial electron energy carried by the photon in the 
laboratory frame. We use the well-known Weizacker-Williams approximation for  
bremsstrahlung photon distribution from the electron:
\begin{gather}
  f_{\gamma/e}(y) = \frac{\alpha}{2\pi}\left(\frac{1+(1-y)^2}{y}\ln\frac{Q^2_{\text{max}}}{Q^2_{\text{min}}}+2m^2_ey\left(\frac{1}{Q^2_{\text{max}}}-\frac{1}{Q^2_{\text{min}}}\right)\right),
  \label{eq:WWA}
\end{gather}
\noindent
where $m_e$ is the electron mass, $Q^2_{\rm min} = m_e^2y^2/(1-y)^2$ and $Q^2_{\rm max} = 1$~GeV$^2$, which is a 
typical value for photoproduction measurements at the HERA. 

As it was already mentioned above,  
direct contribution to the prompt photon photoproduction
comes from the ${\cal O}(\alpha^2)$ deep inelastic Compton scattering subprocess, $\gamma + q \to \gamma + q$.
In the consideration below we decompose the quark content of a proton 
into the sea and valence parts and exploit the idea that the sea quarks all appear
from a perturbative chain as a result of the QCD evolution of gluon density\footnote{A similar treatment has been used recently
in the prompt photon hadroproduction case\cite{Motyka-photon}.}.
Then, appending an explicit gluon splitting vertex, we come to the ${\cal O}(\alpha^2 \alpha_s)$ photon-gluon
fusion subprocess
\begin{gather}
  \gamma(k_1) + g^*(k_2) \to \gamma(p) + q(p_1) + \bar q(p_2).
  \label{eq:vg2vqq}
\end{gather}
\noindent 
Here $q$ denotes any quark flavor and the particles four-momenta are given in parentheses.
The subprocess~(\ref{eq:vg2vqq}) covers also the main part of LO resolved photon contributions.
We employ the $k_T$-factorization approach, thus
effectively taking into account a large piece of higher-order pQCD corrections
via CCFM-evolved TMD gluon density in a proton.
In this way, we keep the non-zero virtuality of the
incoming gluon, $k_2^2 \simeq k_{2T}^2 = - {\mathbf k}_{2T}^2 \neq 0$.
The corresponding off-shell production amplitude 
was calculated in~\cite{LMZ-photo}
and is implemented now in the Monte-Carlo generator \textsc{pegasus}\cite{PEGASUS}.
We note that the subprocess~(\ref{eq:vg2vqq}), being supplemented
with the final state parton radiation, 
covers (at least partly) many other off-shell 
subprocesses --- for example, $\gamma + q^* \to \gamma + q + g$, which 
has been taken into account in our previous calculations\cite{LMZ-photo} separately.
The effects of initial and final state parton emissions (ISR and FSR) can be simulated 
using the standard parton showering algorithms, such as implemented into \textsc{cascade} or \textsc{pythia}.
The remaining contributions from valence quarks, important at large $x$, are taken into account
via simple ${\cal O}(\alpha^2)$ deep inelastic Compton scattering subprocess. 
Of course, here one can safely neglect the off-shellness of valence quarks 
and employ the conventional (collinear) QCD factorization.
Additionally we will take into account ${\cal O}(\alpha^2 \alpha_s^2)$ box contribution:
\begin{gather}
  \gamma(k_1) + g^*(k_2) \to \gamma(p_1) + g(p_2),
  \label{eq:vg2vg}
\end{gather}
\noindent 
since it is known to be sizeable due to high gluon luminosity
at small $x$.
Here we improve the previous consideration\cite{LMZ-photo}
by taking into account non-zero transverse momentum
of initial gluon and use analytical expressions\cite{KNS-photo}
for off-shell production amplitude\footnote{We would like to note that the expression given 
in the Appendix of\cite{KNS-photo} is a bit misleading. Namely, the denominator of (A5) from\cite{KNS-photo} must 
be taken universal for the helicity amplitudes (A1) --- (A3); alternatively, 
one can multiply (A9) by $(u+t_1)/(s+t_1)$.}, where the one-loop integrals were evaluated with \textsc{qcdloop}\cite{QCDLoop}.

To calculate the photon-proton interaction cross section 
in the $k_T$-factorization QCD approach
one has to convolute the corresponding partonic cross section 
with the TMD gluon density in a proton $f_g(x,\mathbf k_{T}^2,\mu^2)$:
\begin{gather}
  d\sigma(\gamma p\to\gamma X) = \int \frac{dx_2}{x_2}\frac{d\phi_2}{2\pi}d\mathbf k_{2T}^2f_a(x_2,\mathbf k_{2T}^2,\mu^2)d\hat\sigma(\gamma g^*\to\gamma X),
  \label{eq:cs}
\end{gather}
\noindent
where $x_2$ and $\phi_2$ are the longitudinal momentum fraction of the colliding proton
carried by incoming off-shell gluon and its azimuthal angle, respectively.
To avoid the collinear divergencies appearing in~(\ref{eq:vg2vqq}) when the final 
photon becomes collinear to the outgoing quark,
we follow an approach\cite{BLZ}
and split the cross section into two pieces:
\begin{gather}
  \sigma = \sigma_{\rm pert}(\mu^2_{\rm reg}) + \sigma_{\rm non-pert}(\mu^2_{\rm reg}), 
\end{gather}
\noindent
where $\sigma_{\rm pert}(\mu^2_{\rm reg})$ is the perturbative contribution
calculated under usual condition when the wavelength of produced photon (in the emitting quark rest frame) 
becomes larger than the typical hadronic scale, ${\cal O}(1 \, {\rm GeV}^{-1})$.
Below this scale, different non-perturbative effects, including photon 
fragmentation, have to be taken into account. 
To separate these two pieces the regularization scale $\mu_{\rm reg}$ is used.
Following\cite{BLZ}, we restrict $\sigma_{\rm pert}(\mu^2_{\rm reg})$ to the 
region $M \geq \mu_{\rm reg}$, where $M$ is the invariant mass of the photon + quark 
subsystem and $\mu_{\rm reg} \sim 1$~GeV is the hadronic scale.
Under this condition, the contribution $\sigma_{\rm pert}(\mu^2_{\rm reg})$ is free from collinear divergences.
Moreover, a special photon isolation criterion (so-called cone isolation), introduced 
in experimental analyses\cite{PromptPhoton-H1, PromptPhoton-ZEUS1, PromptPhoton-ZEUS2}, 
in our calculations is used as a tool\footnote{Another photon isolation criterion was also proposed~\cite{IsolationFrixione}.} to remove
the non-perturbative part $\sigma_{\rm non-pert}(\mu^2_{\rm reg})$, where the final 
photon is radiated close to the quark (inside the isolation cone).
The sensitivity of the numerical results to $\mu_{\rm reg}$ is reasonably soft
if the photon isolation condition is applied (see\cite{BLZ} for more information).
We note also that under this condition the fragmentation contributions 
become negligible\cite{PromptPhoton-H1, PromptPhoton-ZEUS1, PromptPhoton-ZEUS2}, so we can safely omit it in the consideration below.

Concerning the TMD gluon density,
in the present paper we have tested two latest sets obtained 
from the numerical solution of CCFM evolution equation.
First of them is the JH'2013 set 2\cite{JH2013}, which is widely used in phenomenological 
applications (see, for example,\cite{Motyka-photon, LMJ-PP, LM-Higgs} and references therein).
The parameters of the corresponding (rather empirical) initial 
gluon density were derived from a fit to precision 
HERA data on the proton structure functions $F_2(x, Q^2)$ and 
$F_2^c(x, Q^2)$ at $x < 5 \cdot 10^{-3}$ and $Q^2 > 3.5$~GeV$^2$. 
Also we have tried a very recent TMD gluon distribution in a proton, namely, LLM'2022\cite{LLM-2022}.
In contrast with JH'2013 set 2, the analytical expression for the starting 
distribution was chosen in a more physically motivated way --- namely, from the best description of the LHC data on
charged hadron production at low transverse momenta $p_T \sim 1$~GeV in the 
framework of modified soft quark-gluon string model (QGSM)\cite{ModifiedSoftQuarkGluonStringModel-1, ModifiedSoftQuarkGluonStringModel-2}.
Moreover, the gluon saturation effects, which are important at low scales, were taken into account.
Some phenomenological parameters were determined 
from the LHC and HERA data on several hard QCD processes (see\cite{LLM-2022, LLM-2023} for more information). 
Both these TMD gluon densities are available in the Monte-Carlo event generator \textsc{pegasus} and popular \textsc{tmdlib} package\cite{TMDLib2},
which is a C$++$ library providing a framework and an interface to the different parametrizations.

A last important point of our calculations is connected with the proper determination
of the associated jet momentum. In fact, the quarks and/or gluons produced in the hard subprocesses
described above can form final state hadronic jets. However, in addition to that, the produced photon
is accompanied by a number of gluons radiated in the course of the non-collinear evolution,
which also give rise to final jets. 
This is a distinct and remarkable feature of the CCFM-based approach used.
From all of the hadronic jets we choose the one, 
carrying the largest transverse momentum and satisfying the experimental
cuts (leading jet) and then compute the cross section of $\gamma + {\rm jet}$ production.
Technically, we produce a Les Houche Event file\cite{LHE} in the parton level calculations 
performed using the \textsc{pegasus} and 
then process the file with the TMD shower routine implemented into the Monte-Carlo event generator \textsc{cascade}\cite{CASCADE}.
This routine only recently became available for DIS processes, so our work presents the first application of this procedure to photoproduction processes at HERA, to the best of our knowledge.
Thus, we fully reconstruct the CCFM evolution ladder. 
The hadronic jets are reconstructed with the $k_T$-algorithm, implemented in the \textsc{fastjet} tool\cite{FASTJET}.
This method gives us the possibility to take into account the contributions 
from initial state parton showers in a consistent way.
Moreover, it was successfully applied already to investigate 
prompt photon (or Higgs boson) and associated jet production at the LHC (see\cite{LMJ-PP, LM-Higgs}).
Note that additionally we performed final state hadronization procedure, which
is neccesary to
fully reproduce the experimental setup\cite{PromptPhoton-H1, PromptPhoton-ZEUS1, PromptPhoton-ZEUS2}.

\section{Numerical results} \indent

Now we are in a position to present the numerical results. 
First we describe our input and kinematical conditions.
So, as it is often done, we set the renormalization scale 
$\mu_R$ equal to the produced photon transverse energy $E_T^\gamma$.
The factorization scale is defined as $\mu_F^2 = \hat s + {\mathbf Q}^2$
with $\hat s$ and ${\mathbf Q}_T$ being the subprocess invariant energy and net transverse momentum of the initial state, 
respectively. Note that the definition of $\mu_F$ is dictated mainly by the CCFM evolution algorithm (see\cite{CCFM}).
Then, we use the massless 
limit for light ($u$, $d$ and $s$) quarks and set the charm and beauty
masses to $m_c = 1.28$~GeV and $m_b = 4.75$~GeV. 
Also, we apply the two-loop formula for
the strong coupling constant $\alpha_s$ with $N_f = 4$ quark flavours at $\Lambda_{\rm QCD} = 200$~MeV,
as it was originally done in the fits\cite{JH2013,LLM-2022}.

The experimental data for associated prompt photon and jet photoproduction at HERA were
taken by both the H1 and ZEUS collaborations.
The H1 data\cite{PromptPhoton-H1} were obtained in the
following kinematical region\footnote{Here and in the following 
all kinematic quantities are given in the laboratory frame with positive $OZ$ axis
direction given by the proton beam.}: $6 < E_T^\gamma < 15$~GeV, $E_T^{\rm jet} > 4.5$~GeV, $-1.0 < \eta^\gamma < 2.4$,
$-1.3 < \eta^{\rm jet} < 2.3$. The fraction $y$ of
the electron energy transferred to the photon is restricted to the range $0.1 < y < 0.7$. 
More recent ZEUS measurements\cite{PromptPhoton-ZEUS1, PromptPhoton-ZEUS2} refer to the region defined by $6 < E_T^\gamma < 15$~GeV, 
$4 < E_T^{\rm jet} < 35$~GeV, $-0.7 < \eta^\gamma < 0.9$, $ -1.5 < \eta^{\rm jet} < 1.8$ and $0.2 < y < 0.7$.
The data\cite{PromptPhoton-H1, PromptPhoton-ZEUS1, PromptPhoton-ZEUS2} were obtained with the electron energy $E_e = 27.6$~GeV and 
proton energy $E_p = 920$~GeV.

\begin{figure}
\begin{center}
\includegraphics[width=7.0cm]{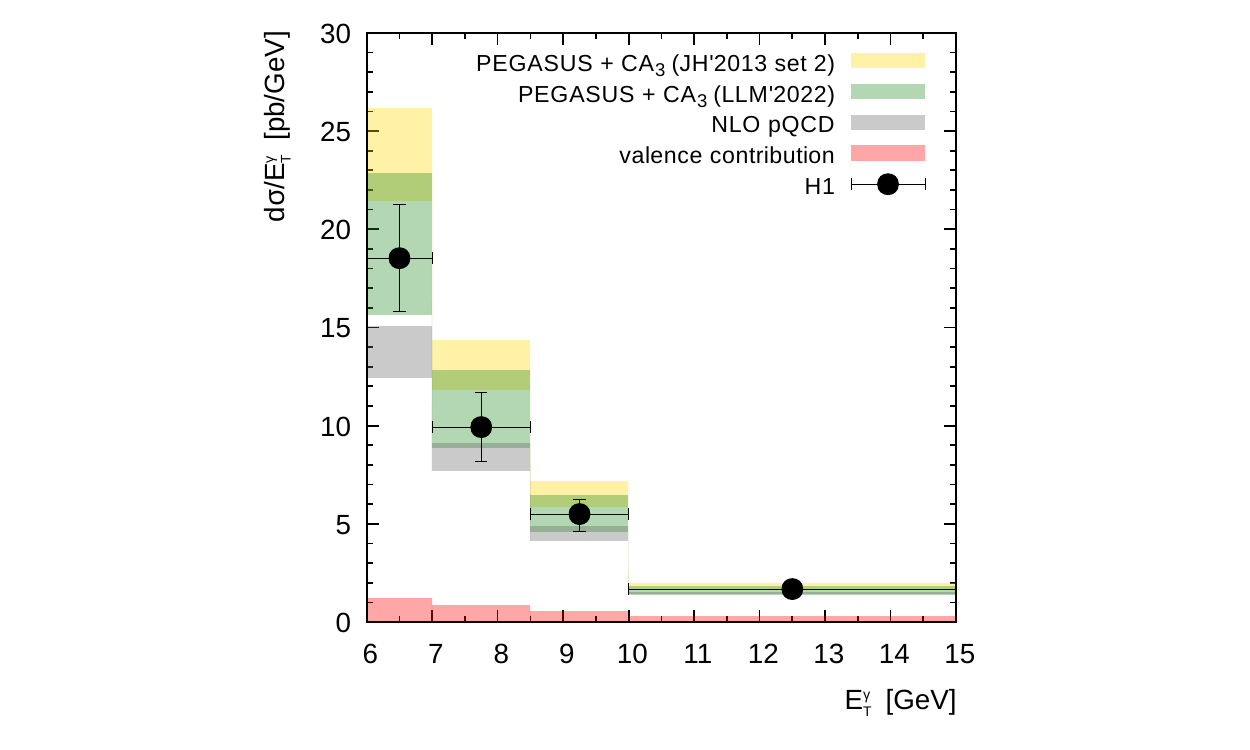} \hspace{0.5cm}
\includegraphics[width=7.0cm]{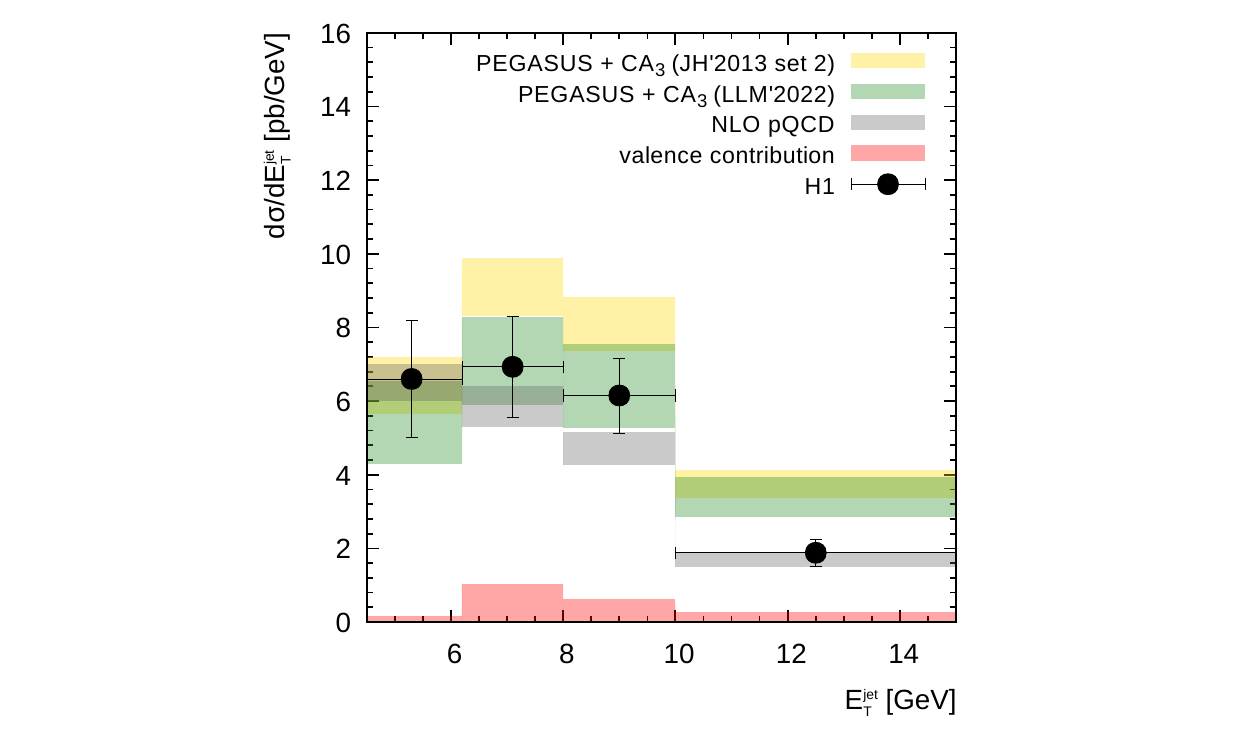}
\includegraphics[width=7.0cm]{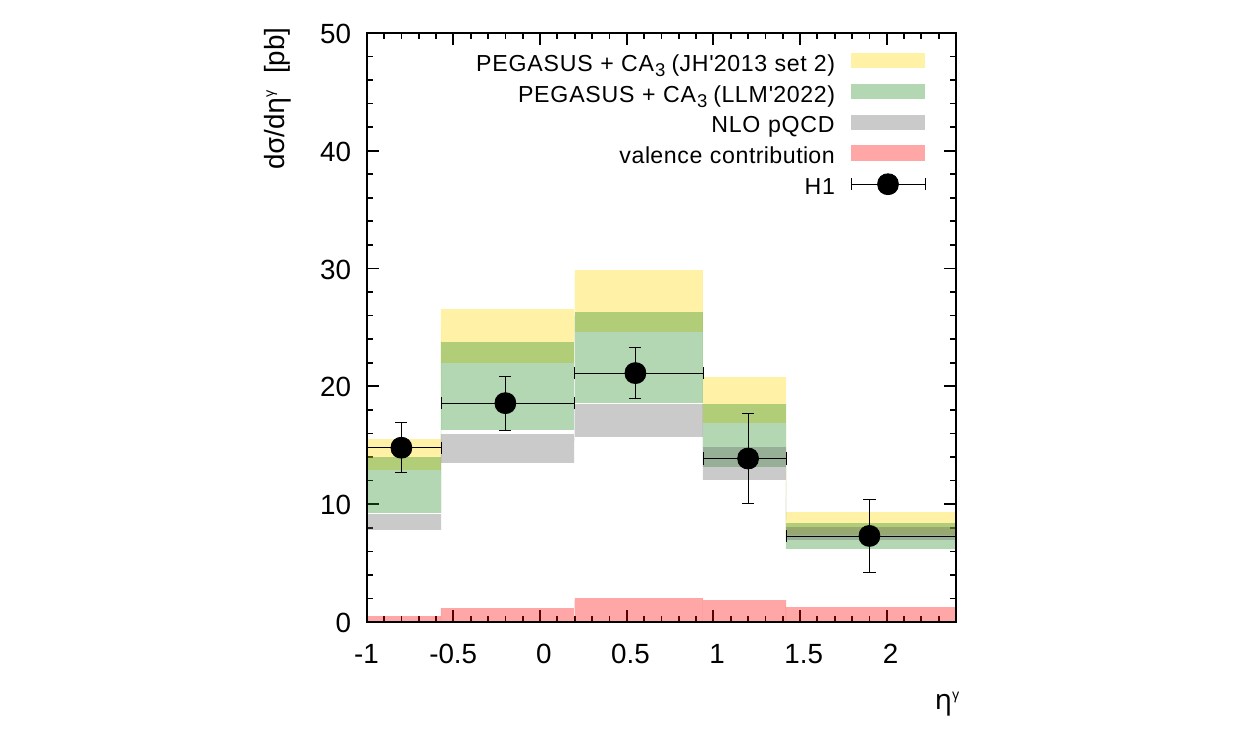} \hspace{0.5cm}
\includegraphics[width=7.0cm]{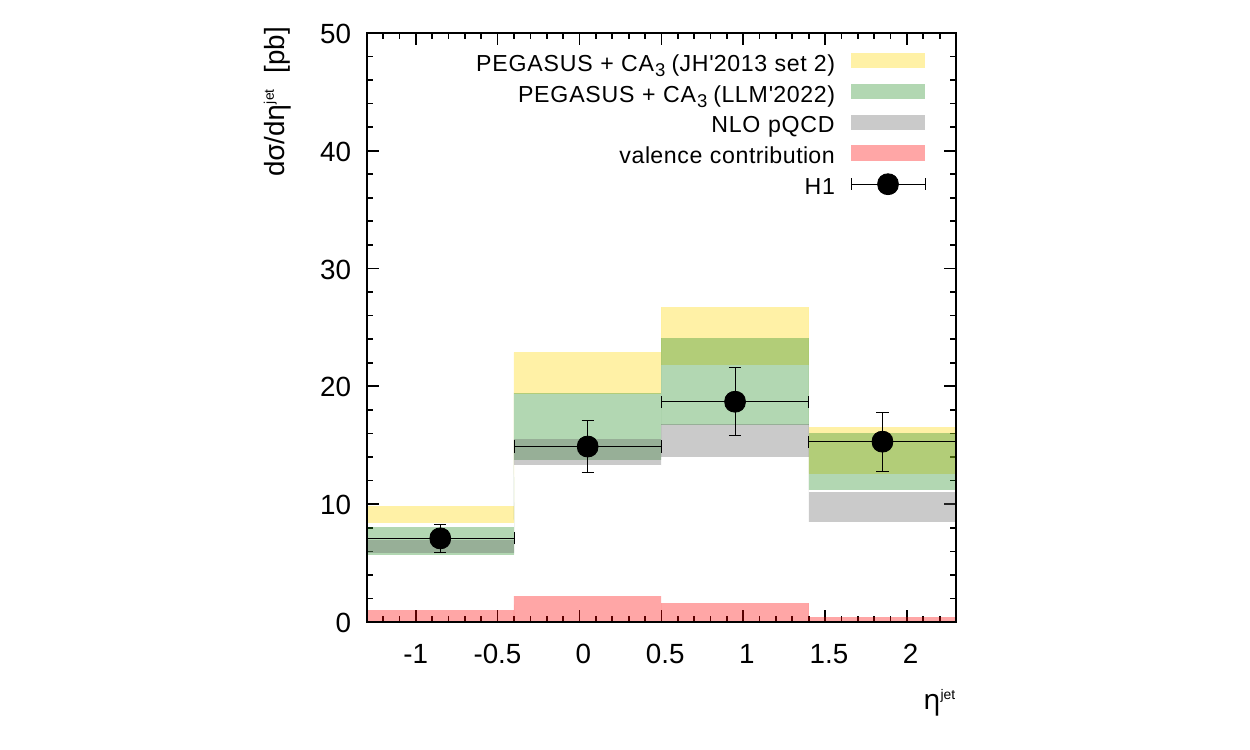}
\caption{The associated prompt photon and jet photoproduction cross
section as functions of photon and jet transverse energies and pseudorapidities. 
The green histograms and shaded bands correspond to the predictions obtained
with LLM'2022 gluon density and estimated scale uncertainties of these calculations.
The yellow histograms represent the JH'2013 set 2 predictions.
Separately shown the contributions from
$\gamma + q_{\rm val} \to \gamma + q$ subprocess and
conventional NLO pQCD results (taken from\cite{PromptPhoton-H1}). 
The experimental data are from H1\cite{PromptPhoton-H1}.}
\label{fig-H1} 
\end{center}
\end{figure}

\begin{figure}
\begin{center}
\includegraphics[width=7.0cm]{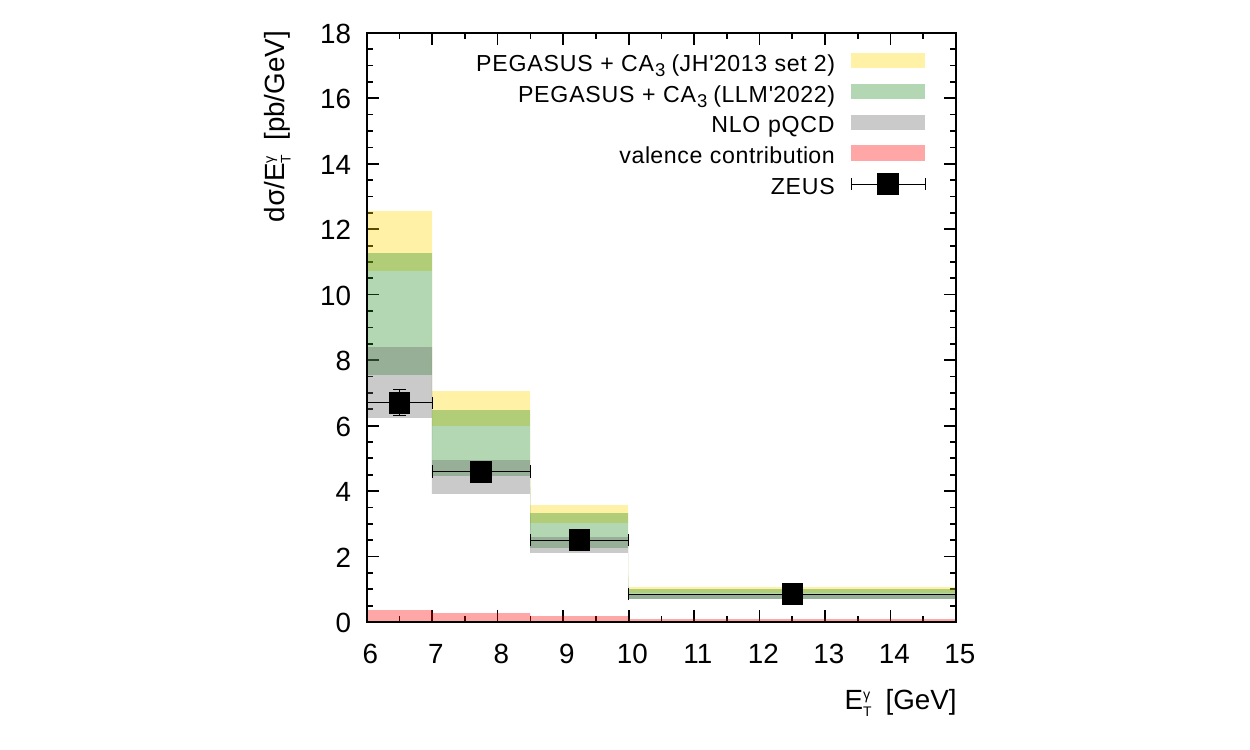} \hspace{0.5cm}
\includegraphics[width=7.0cm]{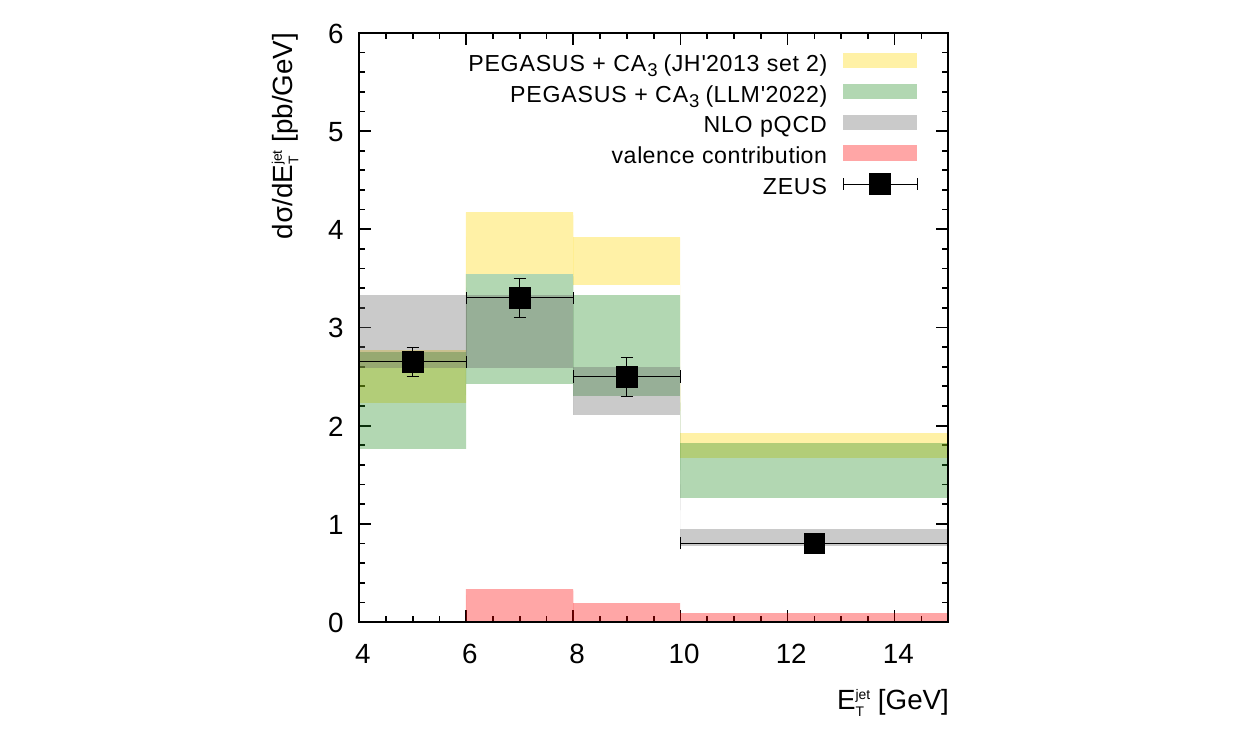}
\includegraphics[width=7.0cm]{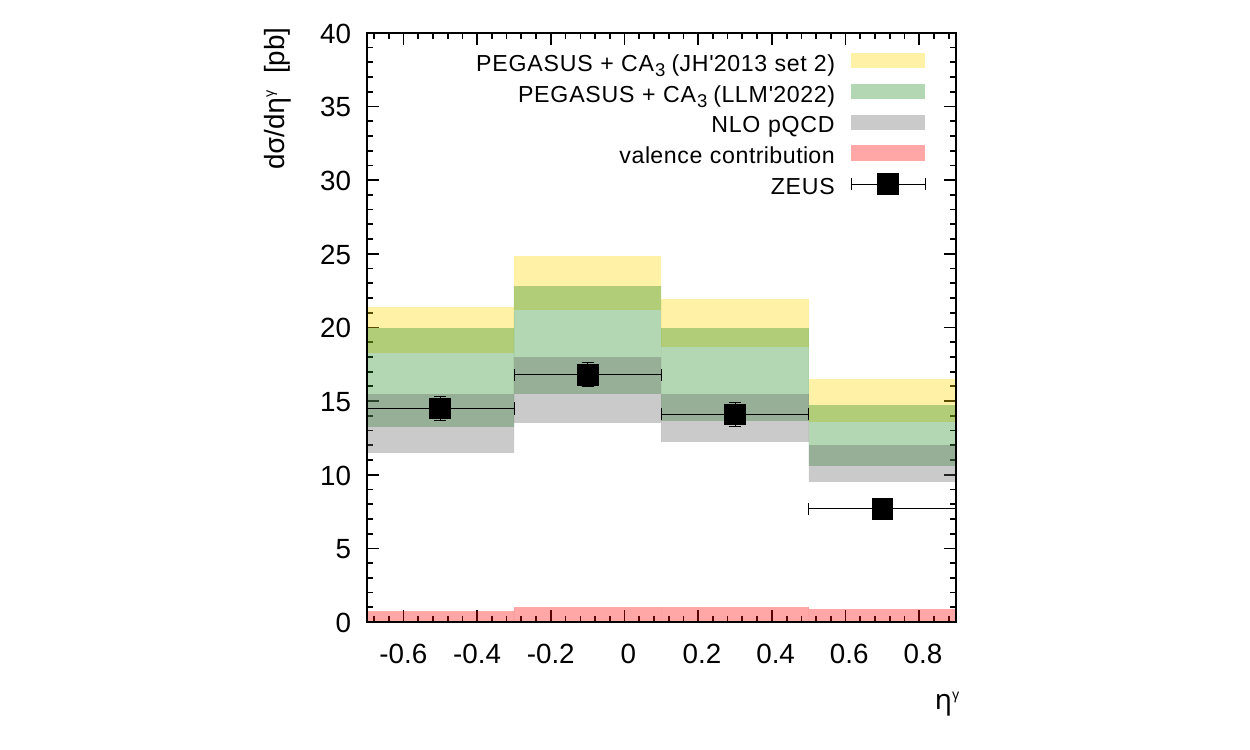} \hspace{0.5cm}
\includegraphics[width=7.0cm]{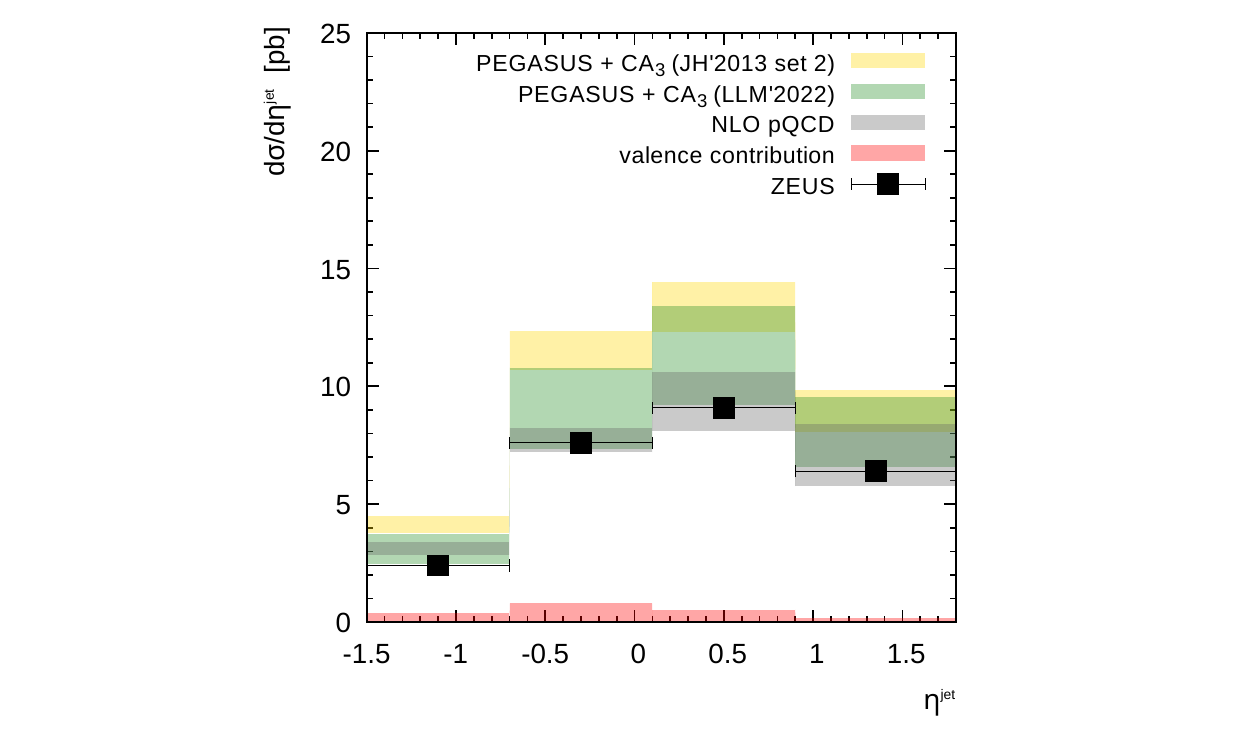}
\includegraphics[width=7.0cm]{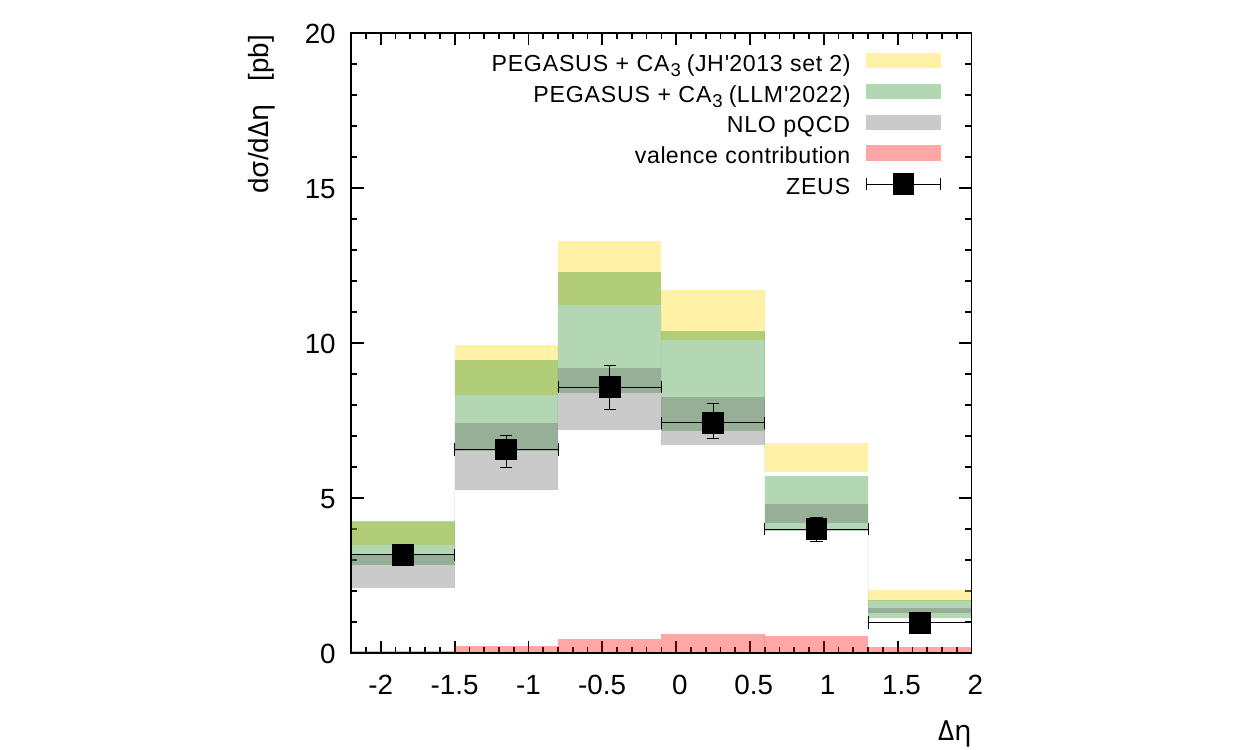} \hspace{0.5cm}
\includegraphics[width=7.0cm]{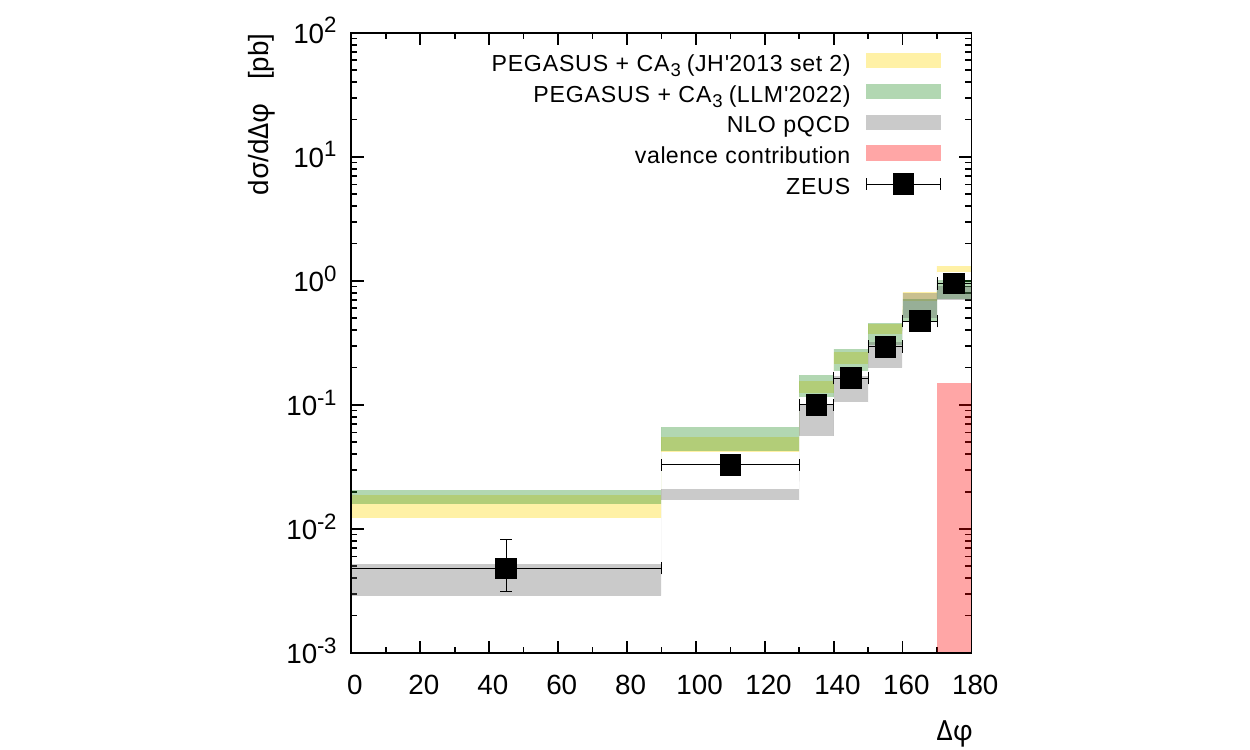}
\caption{The associated prompt photon and jet photoproduction cross
section as functions of photon and jet transverse energies, pseudorapidities, differences in their azimuthal 
angles and pseudorapidities. 
The notations are the same as in Fig.~\ref{fig-H1}.
The experimental data are from ZEUS\cite{PromptPhoton-ZEUS1,PromptPhoton-ZEUS2}.}
\label{fig-ZEUS} 
\end{center}
\end{figure}

\begin{figure}
\begin{center}
\includegraphics[width=7.0cm]{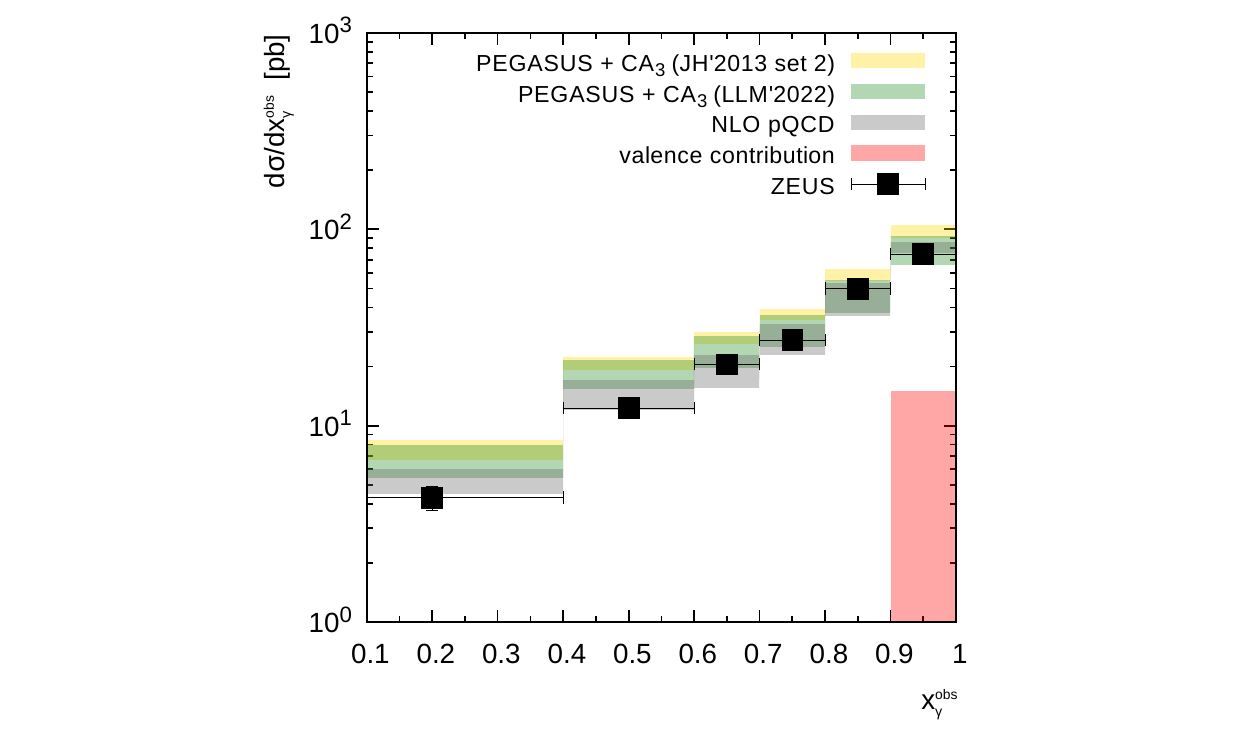} \hspace{0.5cm}
\includegraphics[width=7.0cm]{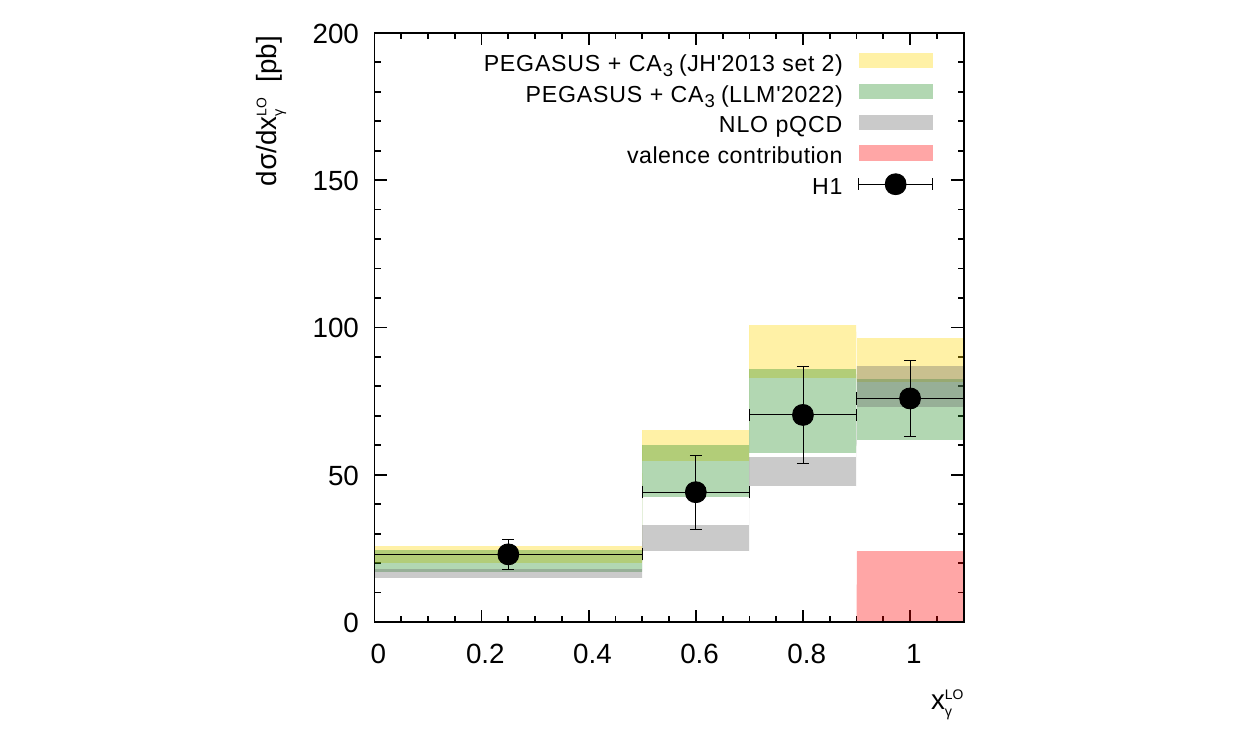}
\includegraphics[width=7.0cm]{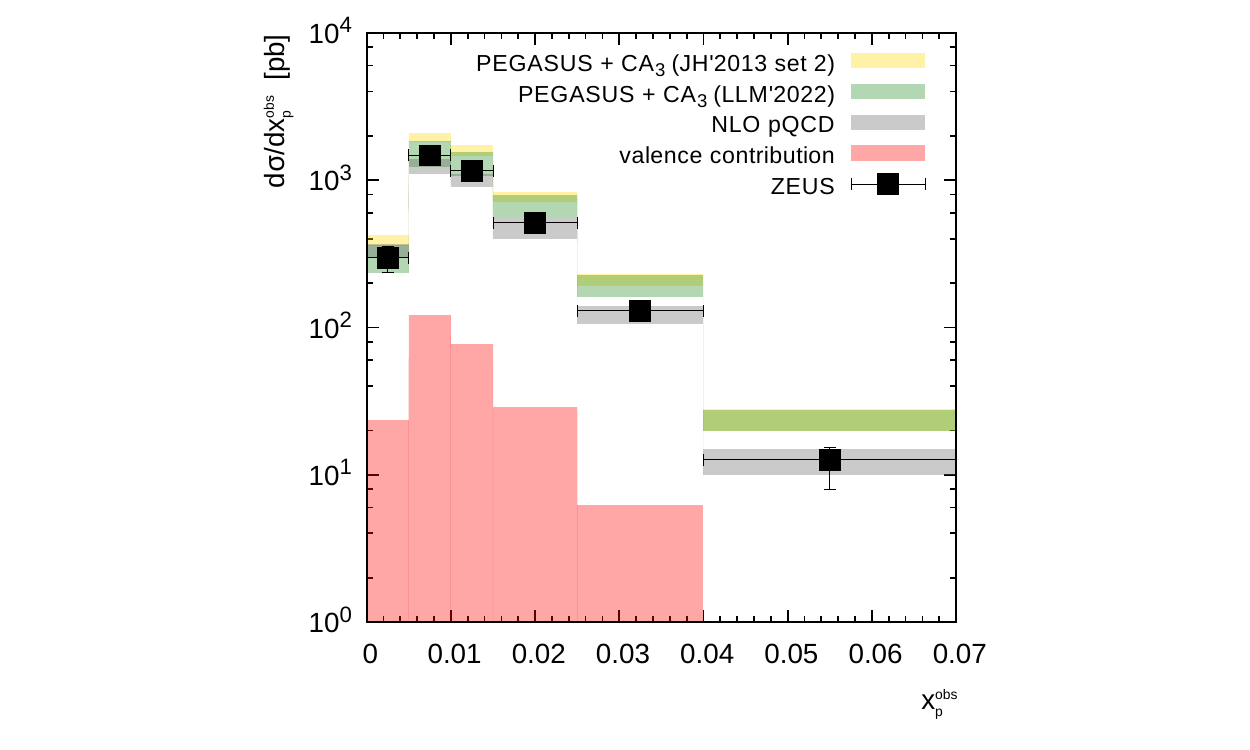} \hspace{0.5cm}
\includegraphics[width=7.0cm]{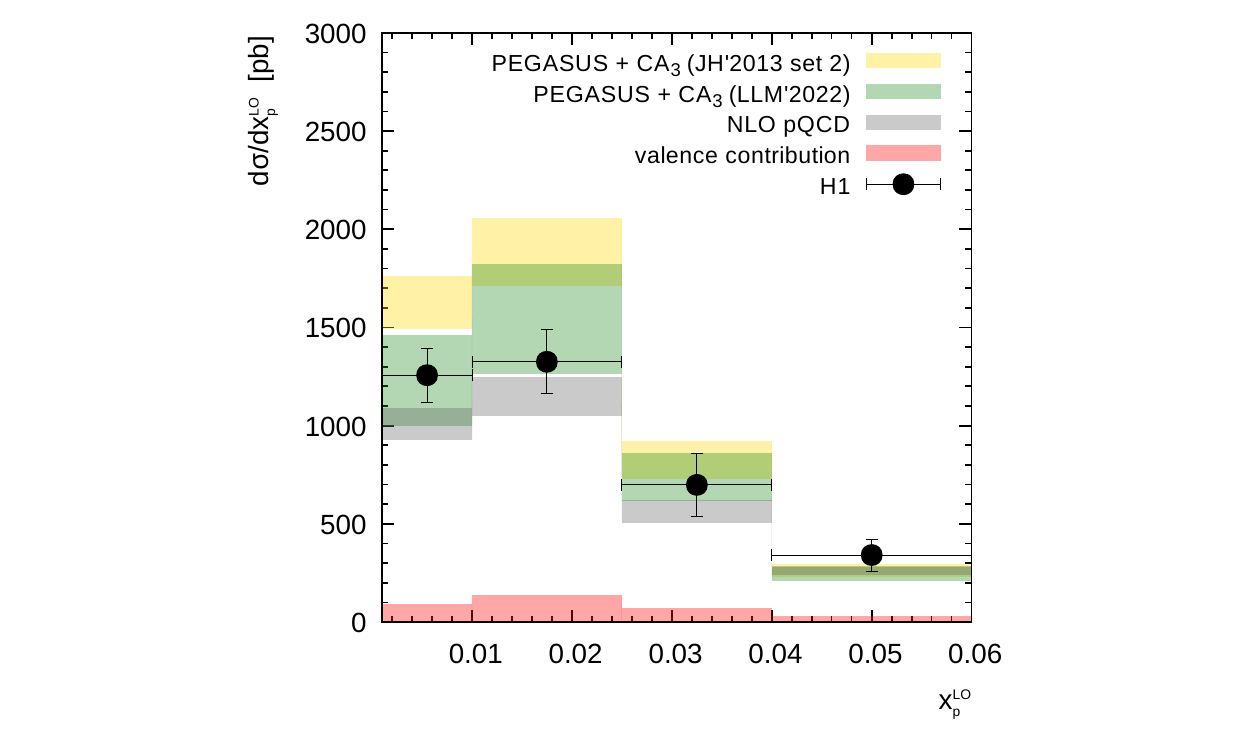}
\caption{The associated prompt photon and jet photoproduction cross
section as functions of $x_\gamma^{\rm LO}$, $x_\gamma^{\rm obs}$, $x_p^{\rm LO}$ and $x_p^{\rm obs}$ variables. 
The notations are the same as in Fig.~\ref{fig-H1}. 
The experimental data are from H1\cite{PromptPhoton-H1} and ZEUS\cite{PromptPhoton-ZEUS1}.}
\label{fig-x} 
\end{center}
\end{figure}

\begin{figure}
\begin{center}
\includegraphics[width=7.9cm]{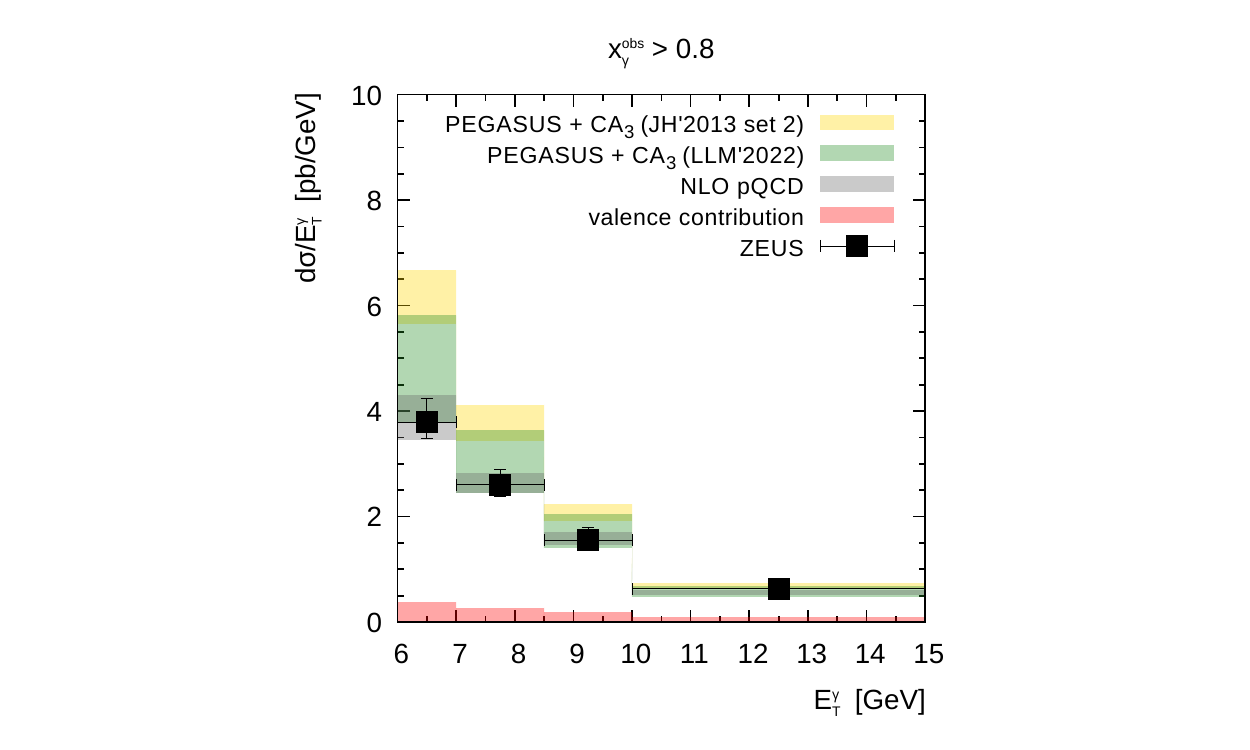}
\includegraphics[width=7.9cm]{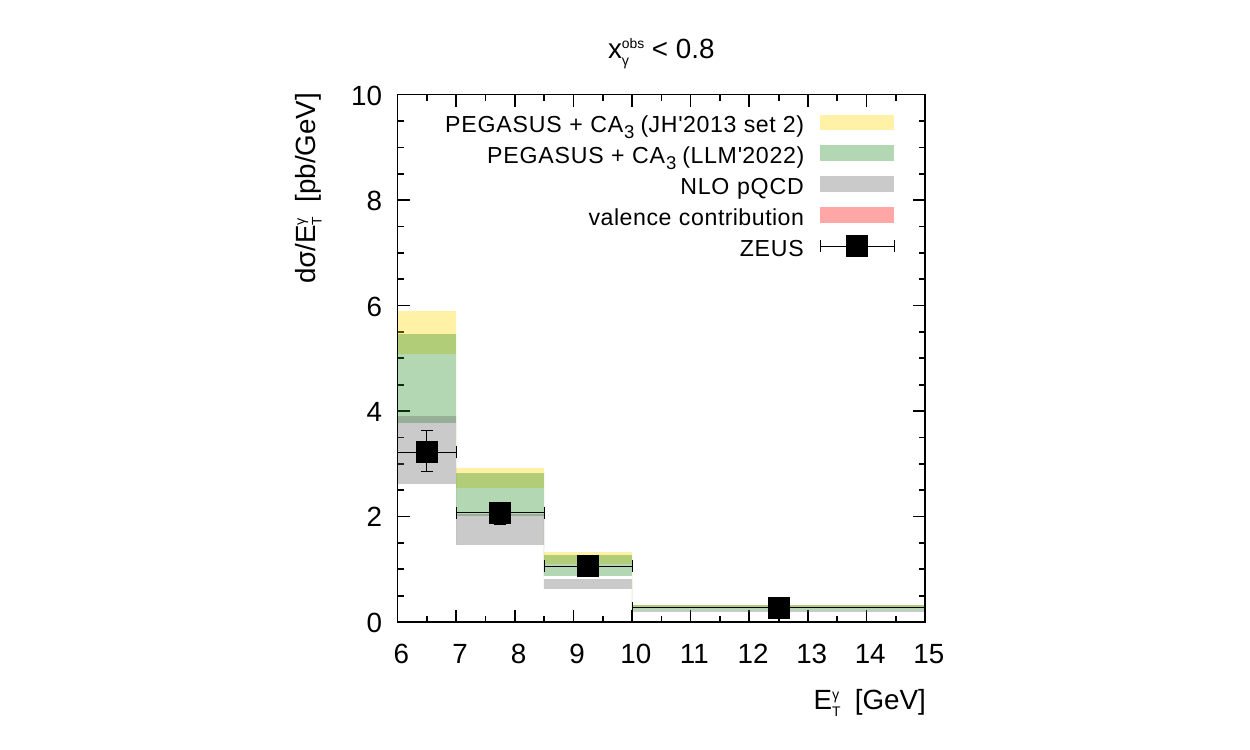}
\includegraphics[width=7.9cm]{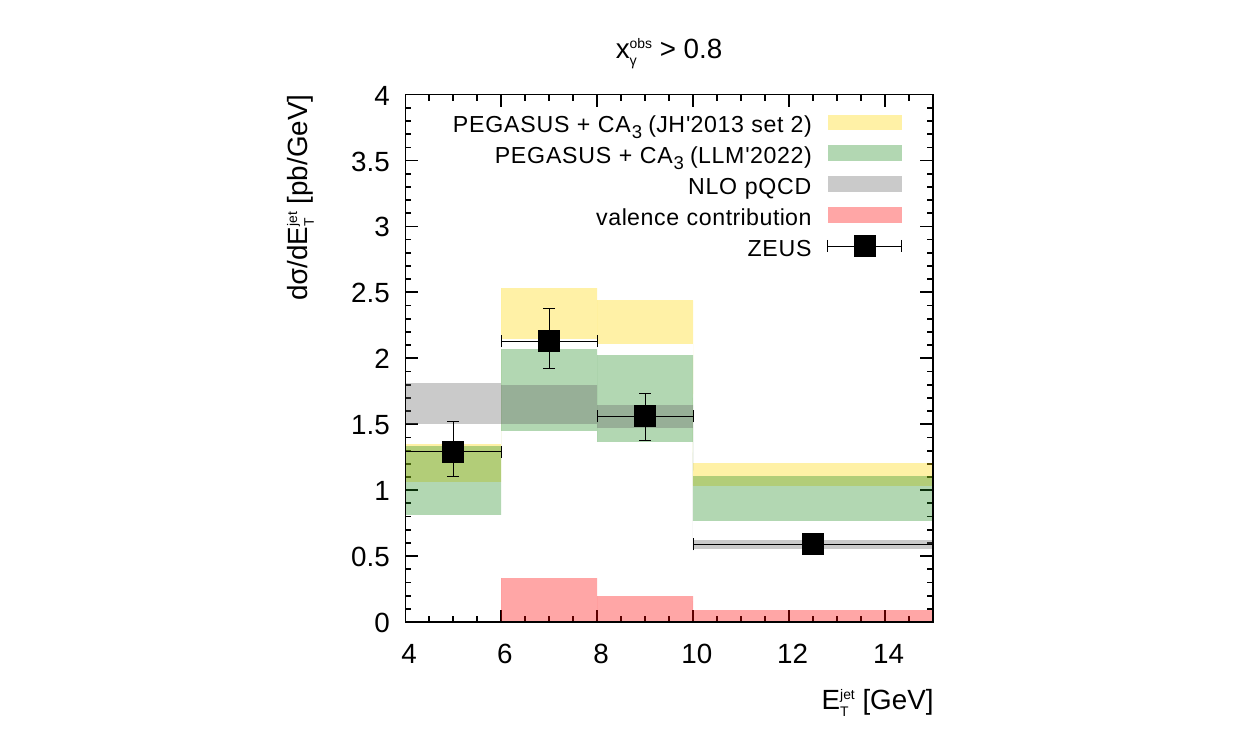}
\includegraphics[width=7.9cm]{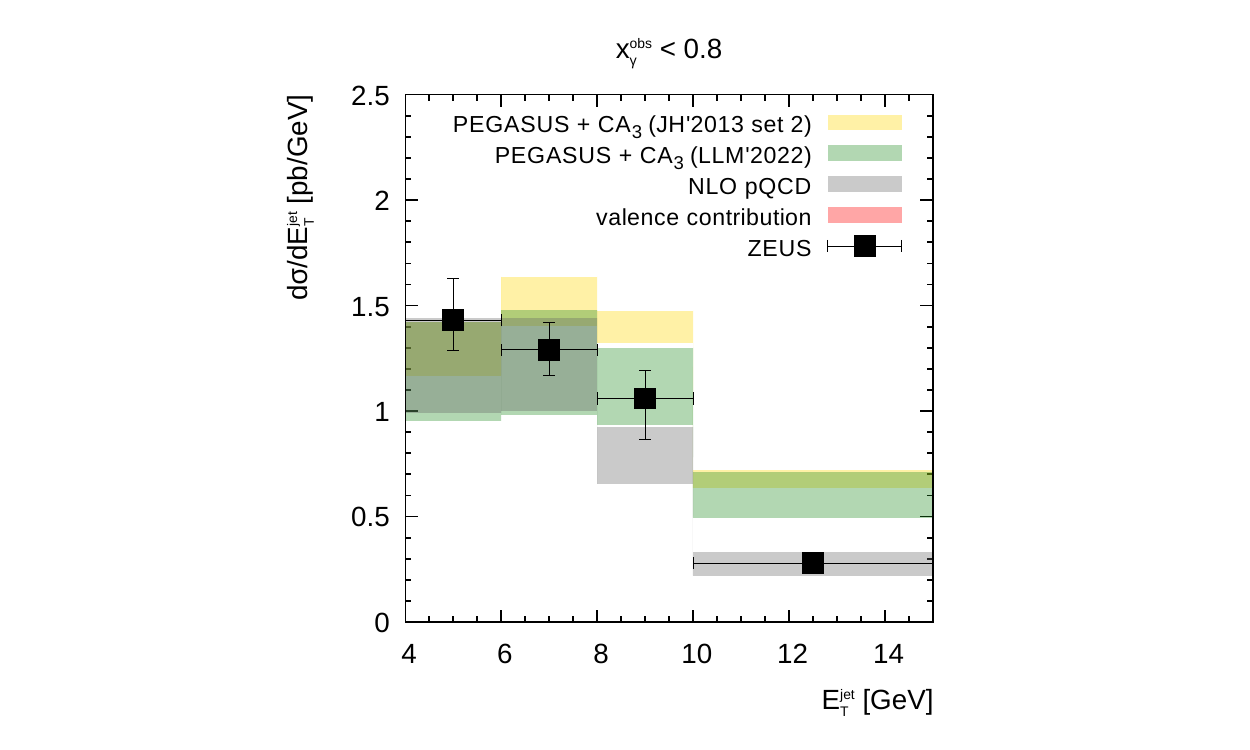}
\caption{The associated prompt photon and jet photoproduction cross
section as functions of photon and jet transverse energies $E_T^\gamma$ and $E_T^{\rm jet}$ 
at $x_\gamma^{\rm obs} > 0.8$ (left panels) and $x_\gamma^{\rm obs} < 0.8$ (right panels). 
The notations are the same as in Fig.~\ref{fig-H1}. The experimental data are from ZEUS~\cite{PromptPhoton-ZEUS2}.}
\label{fig-ZEUS2-pt} 
\end{center}
\end{figure}

\begin{figure}
\begin{center}
\includegraphics[width=7.9cm]{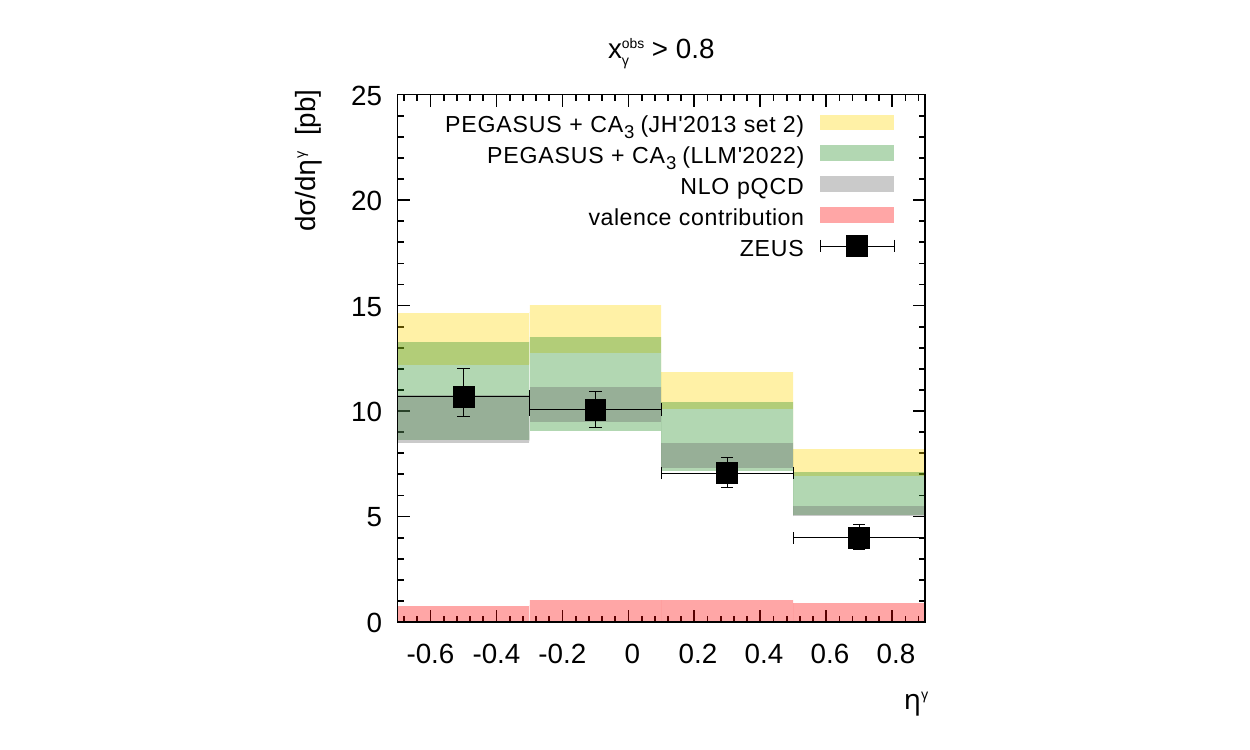}
\includegraphics[width=7.9cm]{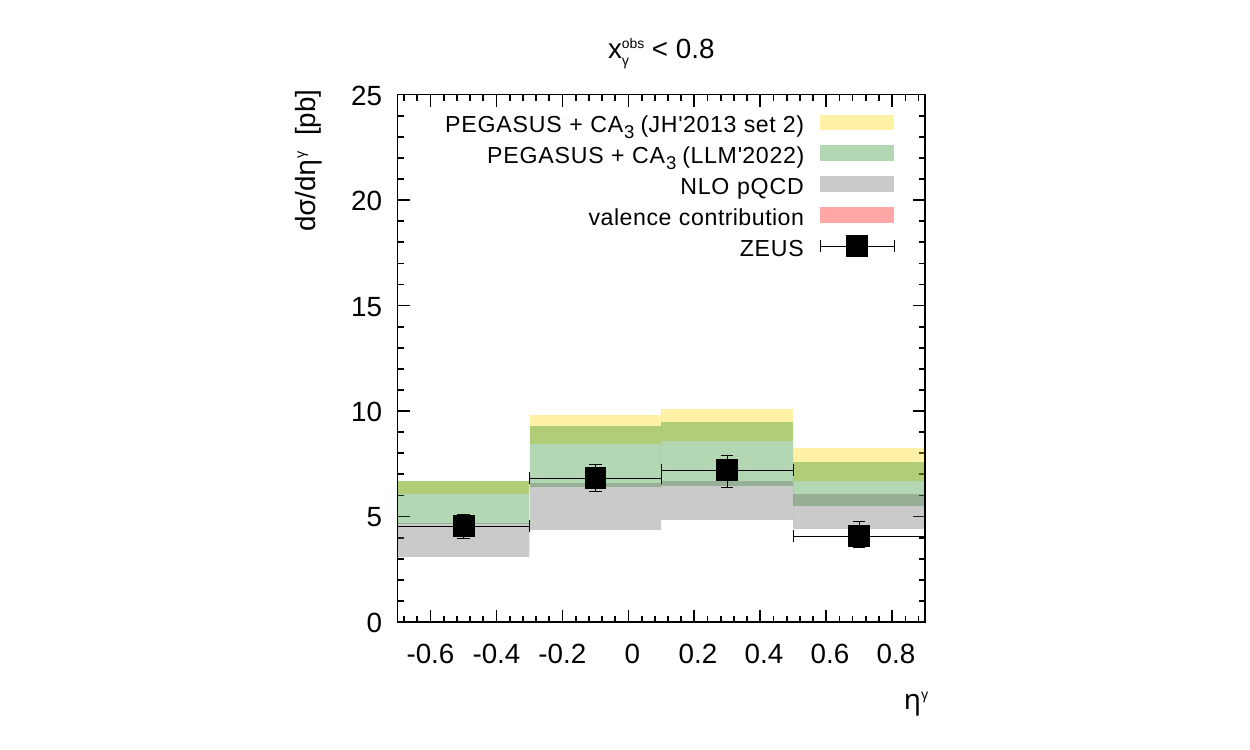}
\includegraphics[width=7.9cm]{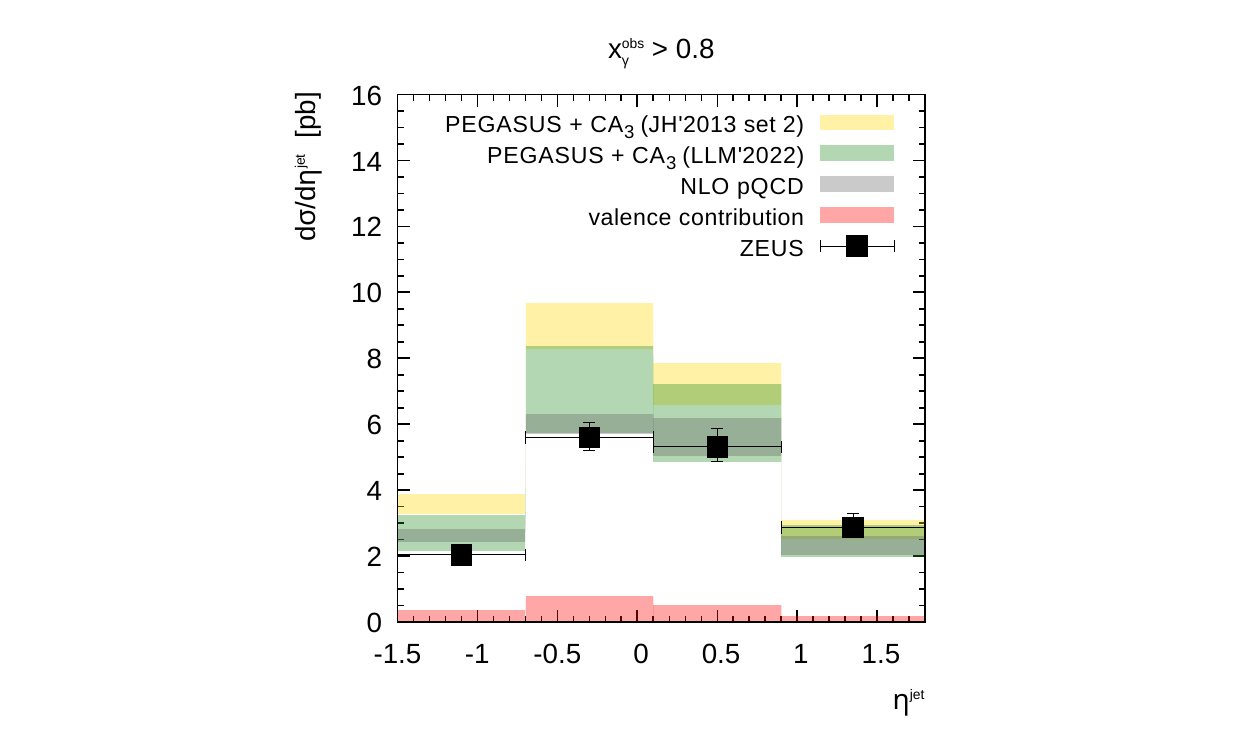}
\includegraphics[width=7.9cm]{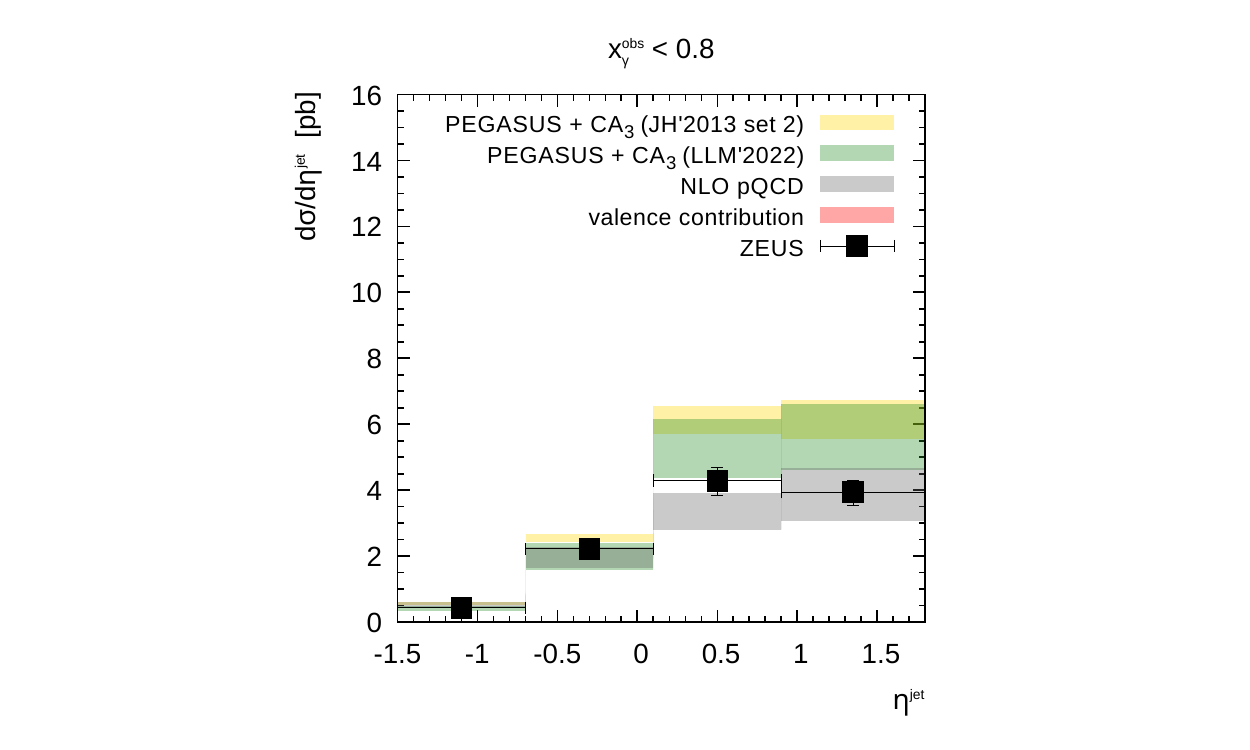}
\caption{The associated prompt photon and jet photoproduction cross
section as functions of photon and jet pseudo-rapidities $\eta^\gamma$ and $\eta^{\rm jet}$ 
at $x_\gamma^{\rm obs} > 0.8$ (left panels) and $x_\gamma^{\rm obs} < 0.8$ (right panels). 
The notations are the same as in Fig.~\ref{fig-H1}. The experimental data are from ZEUS~\cite{PromptPhoton-ZEUS2}.}
\label{fig-ZEUS2-eta} 
\end{center}
\end{figure}

\begin{figure}
\begin{center}
\includegraphics[width=7.9cm]{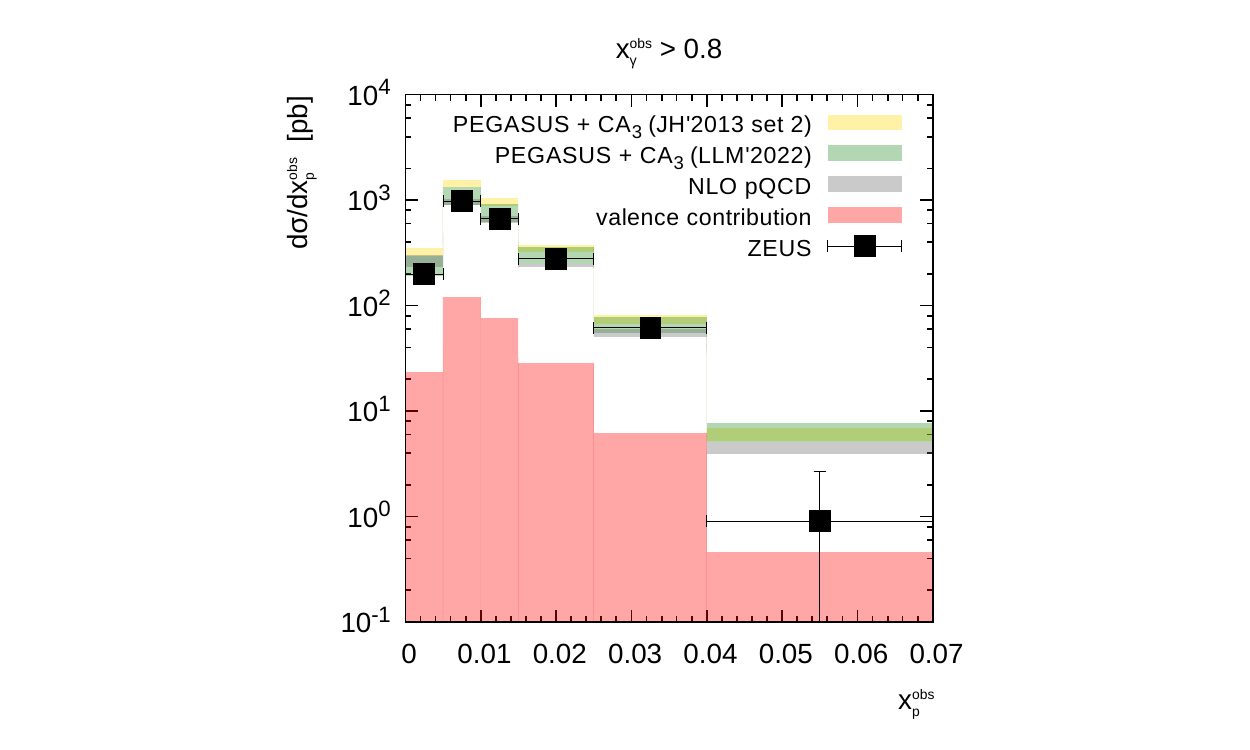}
\includegraphics[width=7.9cm]{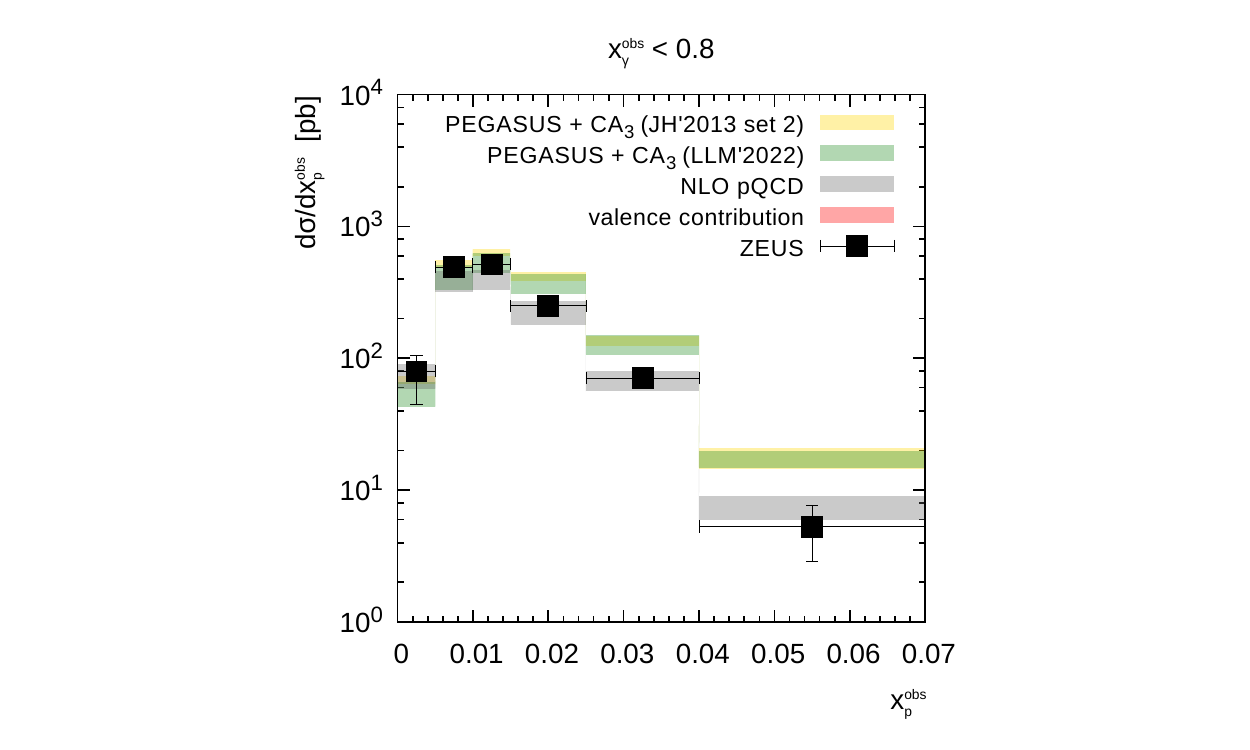}
\caption{The associated prompt photon and jet photoproduction cross
section as functions of $x_p^{\rm obs}$ at $x_\gamma^{\rm obs} > 0.8$ (left panels) and $x_\gamma^{\rm obs} < 0.8$ (right panels). 
The notations are the same as in Fig.~\ref{fig-H1}. The experimental data are from ZEUS~\cite{PromptPhoton-ZEUS2}.}
\label{fig-ZEUS2-oth2} 
\end{center}
\end{figure}

\begin{figure}
\begin{center}
\includegraphics[width=7.9cm]{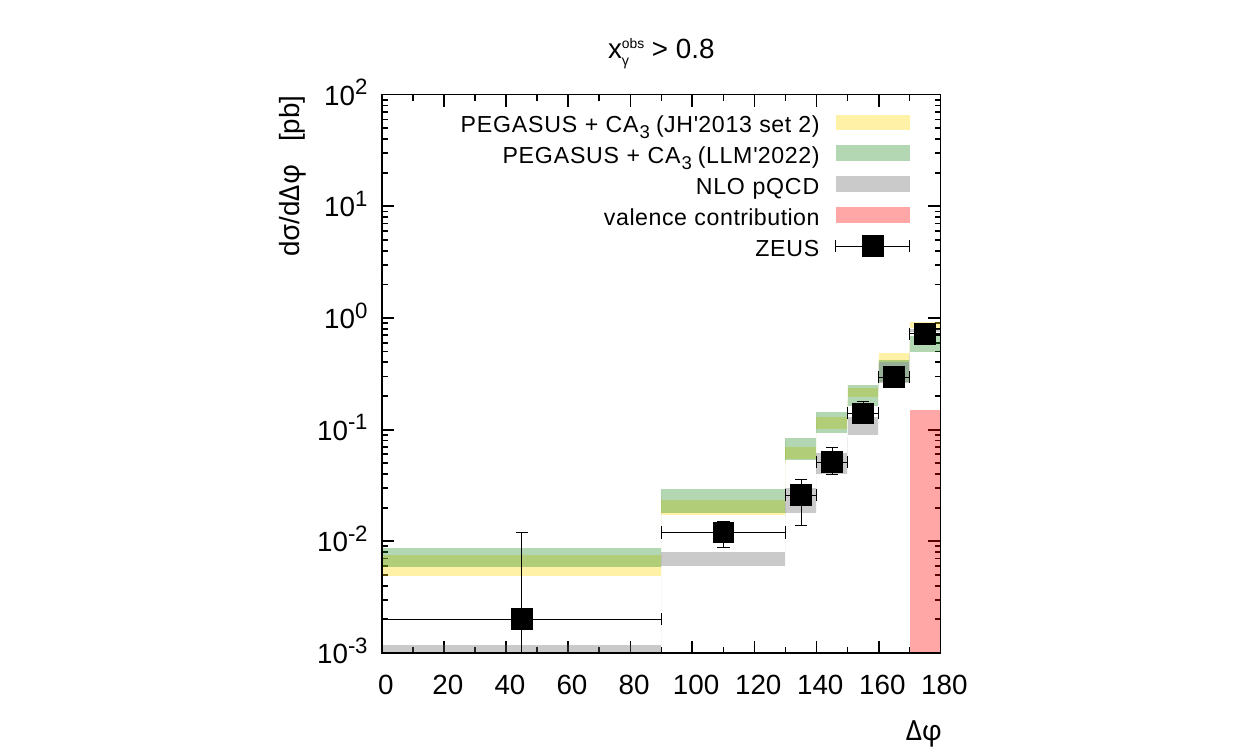}
\includegraphics[width=7.9cm]{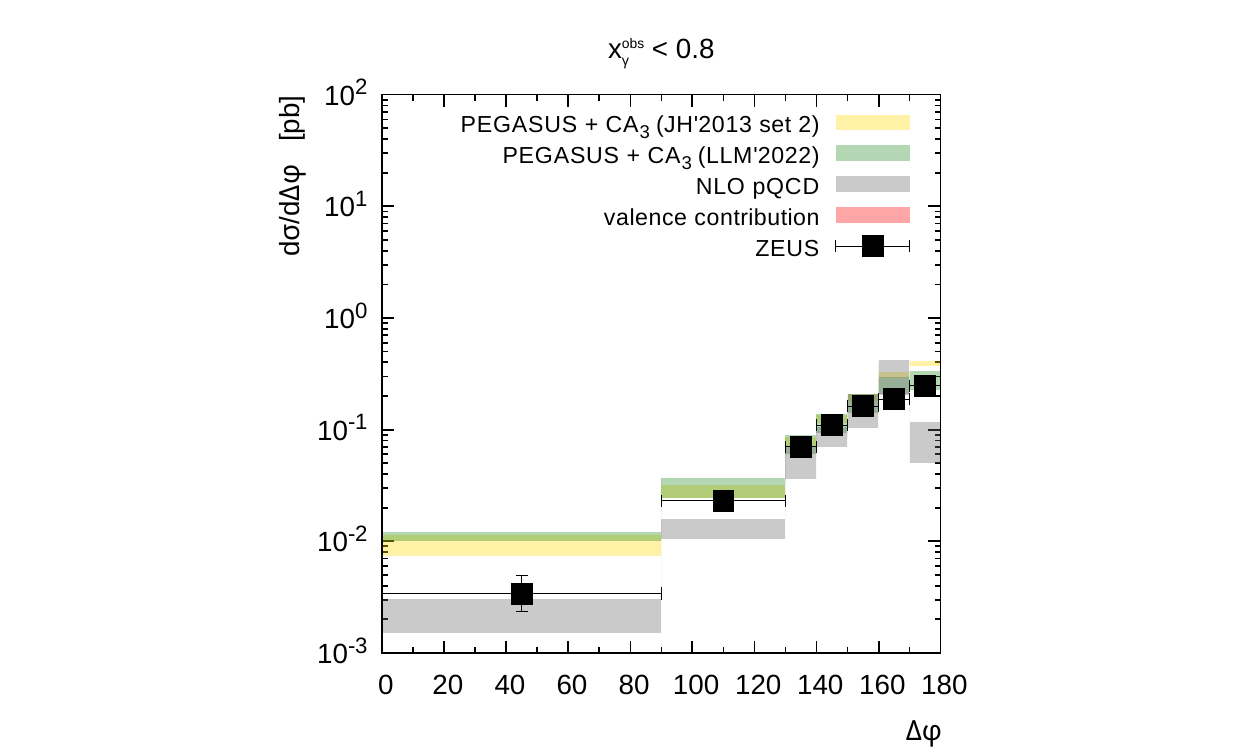}
\includegraphics[width=7.9cm]{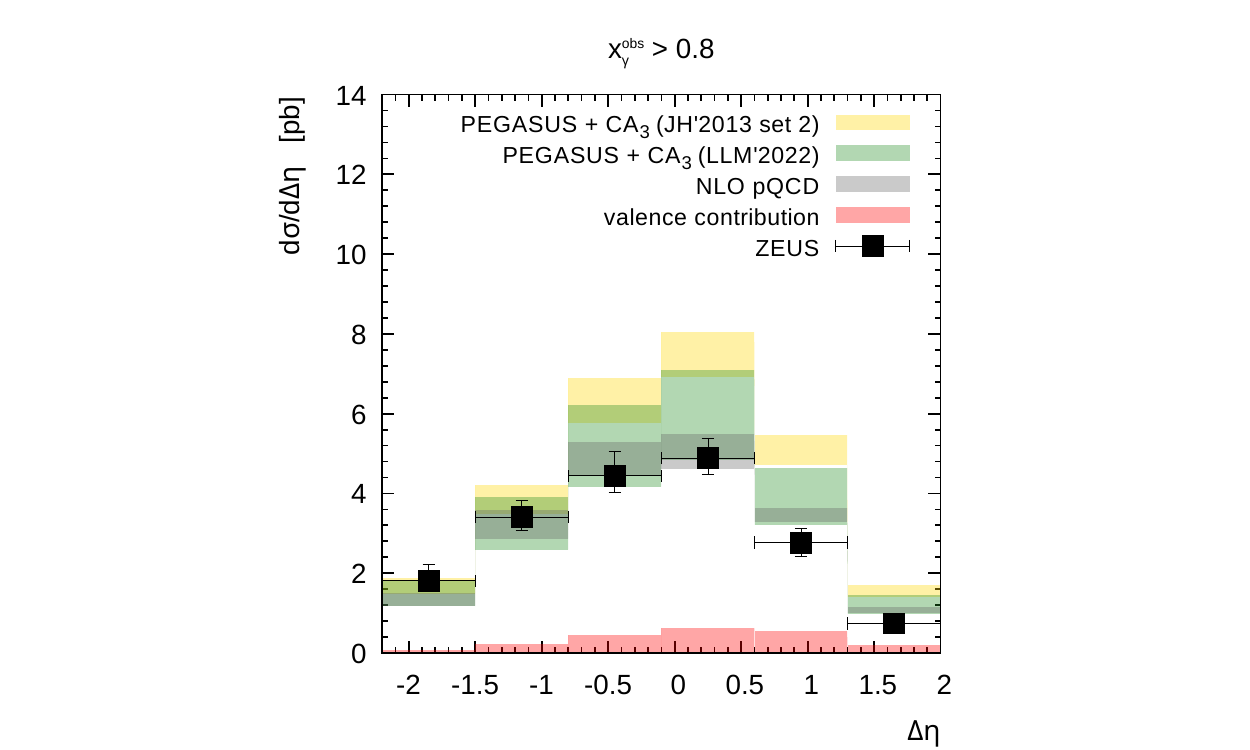}
\includegraphics[width=7.9cm]{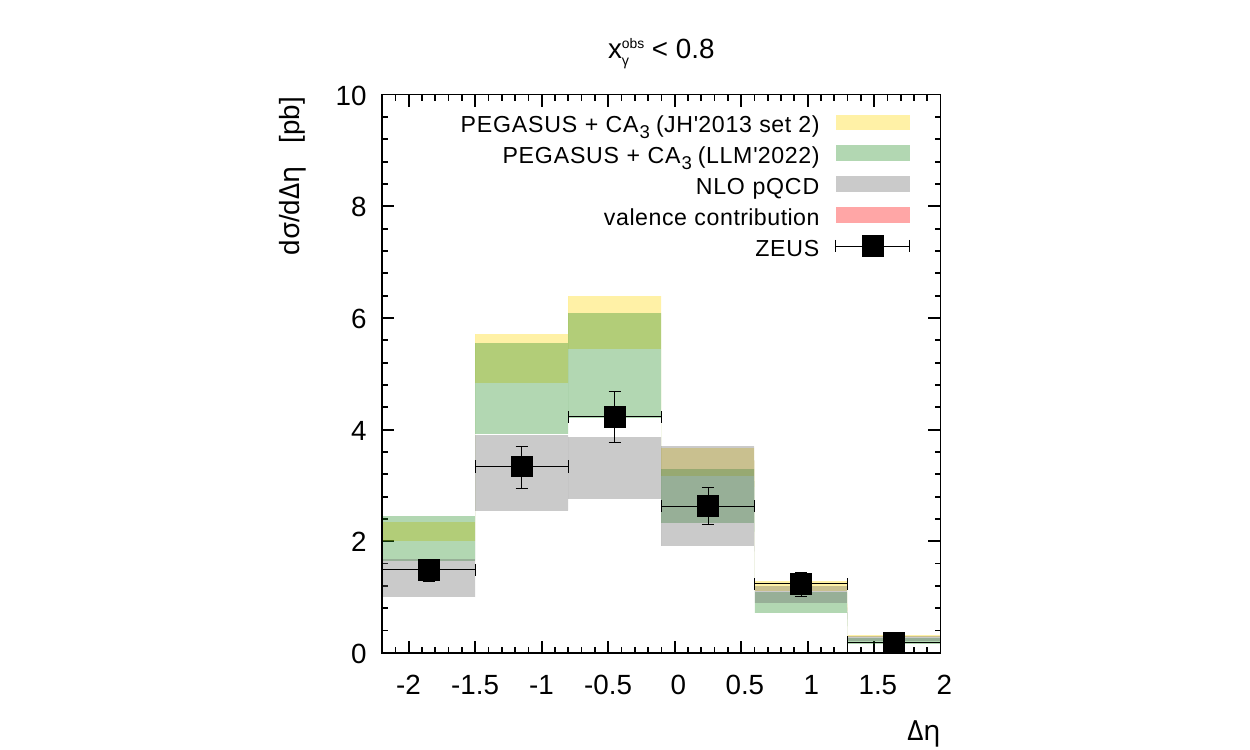}
\caption{The associated prompt photon and jet photoproduction cross
section as functions of the prompt photon and jet azimuthal angle 
and pseudorapidity differences $\Delta\phi$ and $\Delta\eta$ 
at $x_\gamma^{\rm obs} > 0.8$ (left panels) and $x_\gamma^{\rm obs} < 0.8$ (right panels). 
The notations are the same as in Fig.~\ref{fig-H1}. The experimental data are from ZEUS~\cite{PromptPhoton-ZEUS2}.}
\label{fig-ZEUS2-oth} 
\end{center}
\end{figure}

\begin{figure}
\begin{center}
\includegraphics[width=7.9cm]{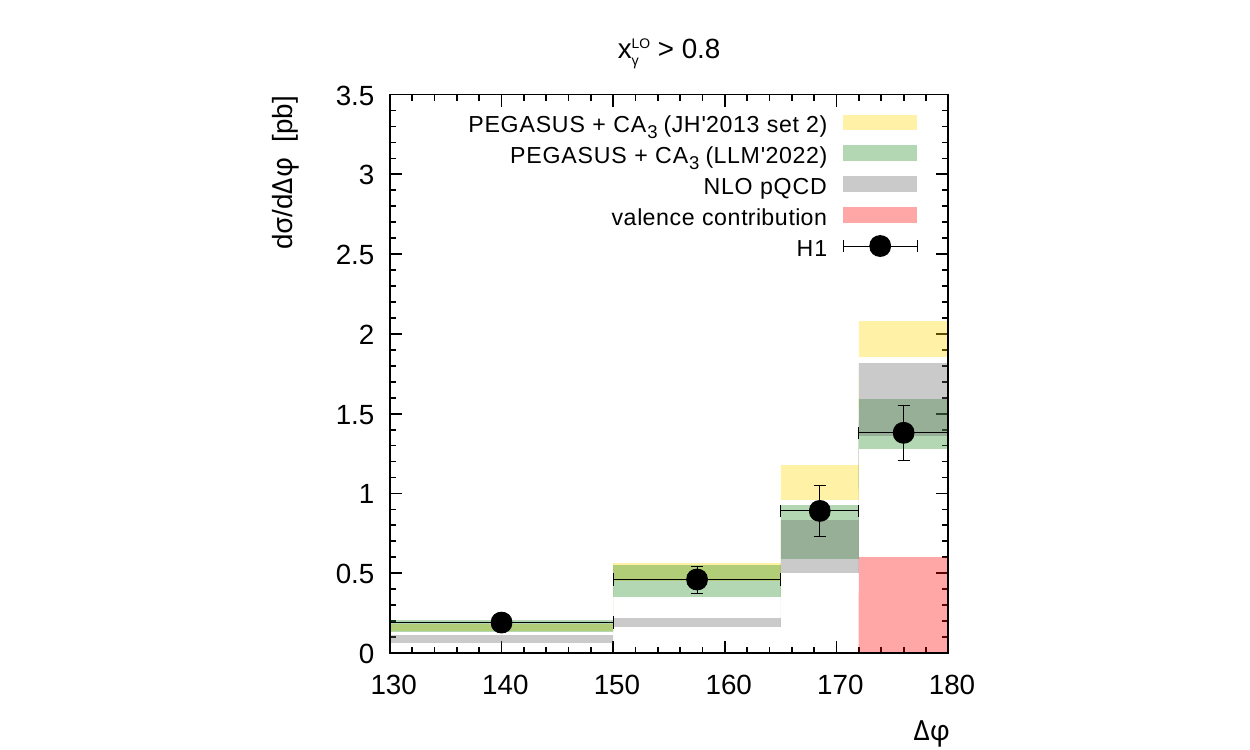}
\includegraphics[width=7.9cm]{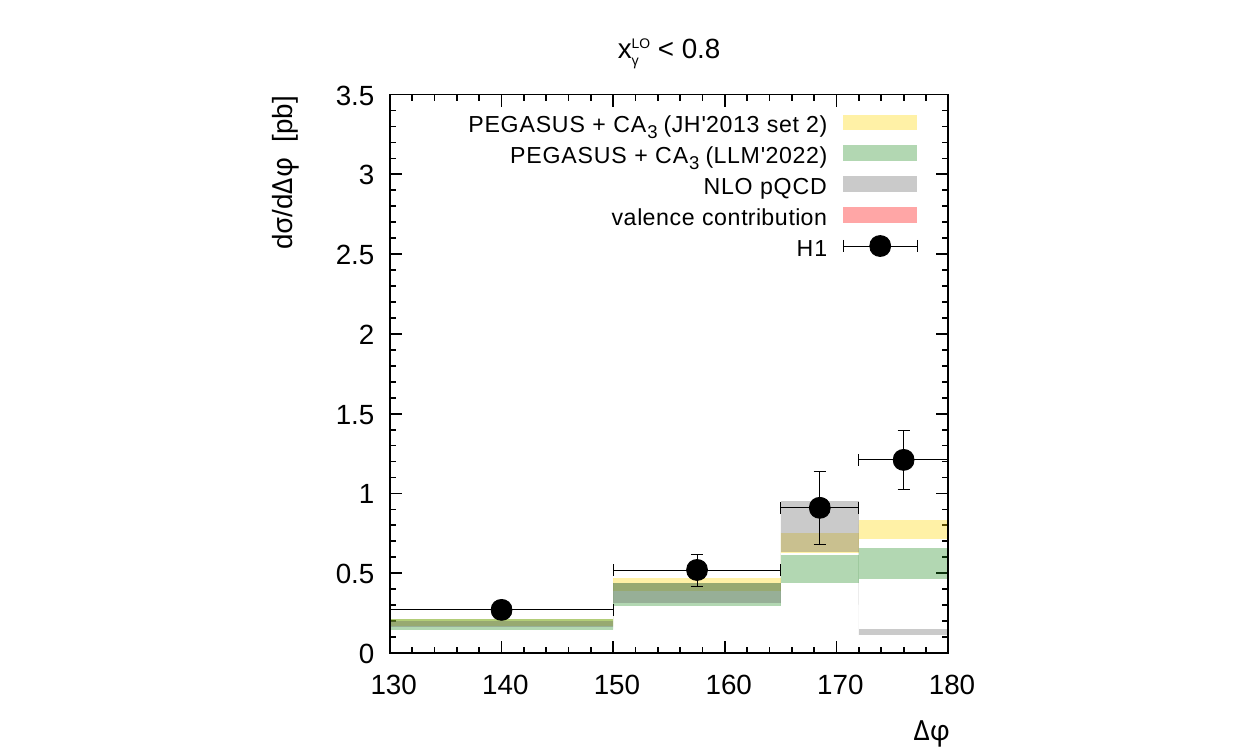}
\includegraphics[width=7.9cm]{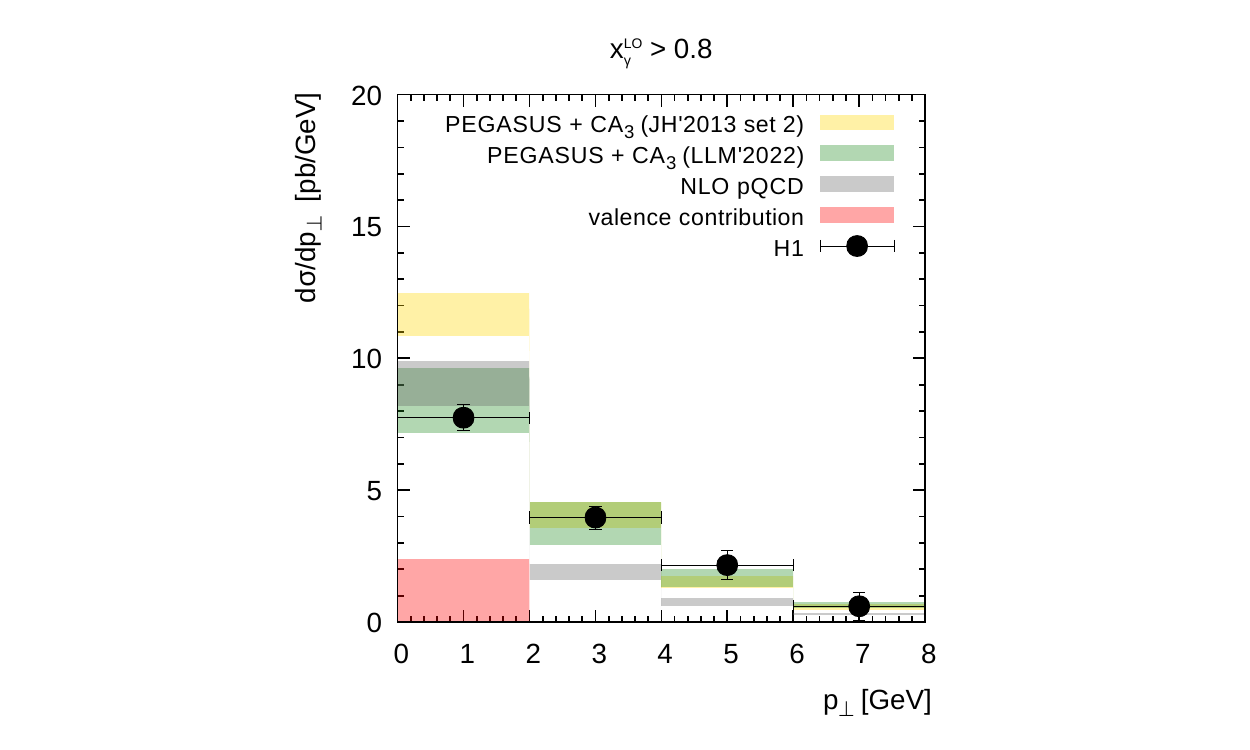}
\includegraphics[width=7.9cm]{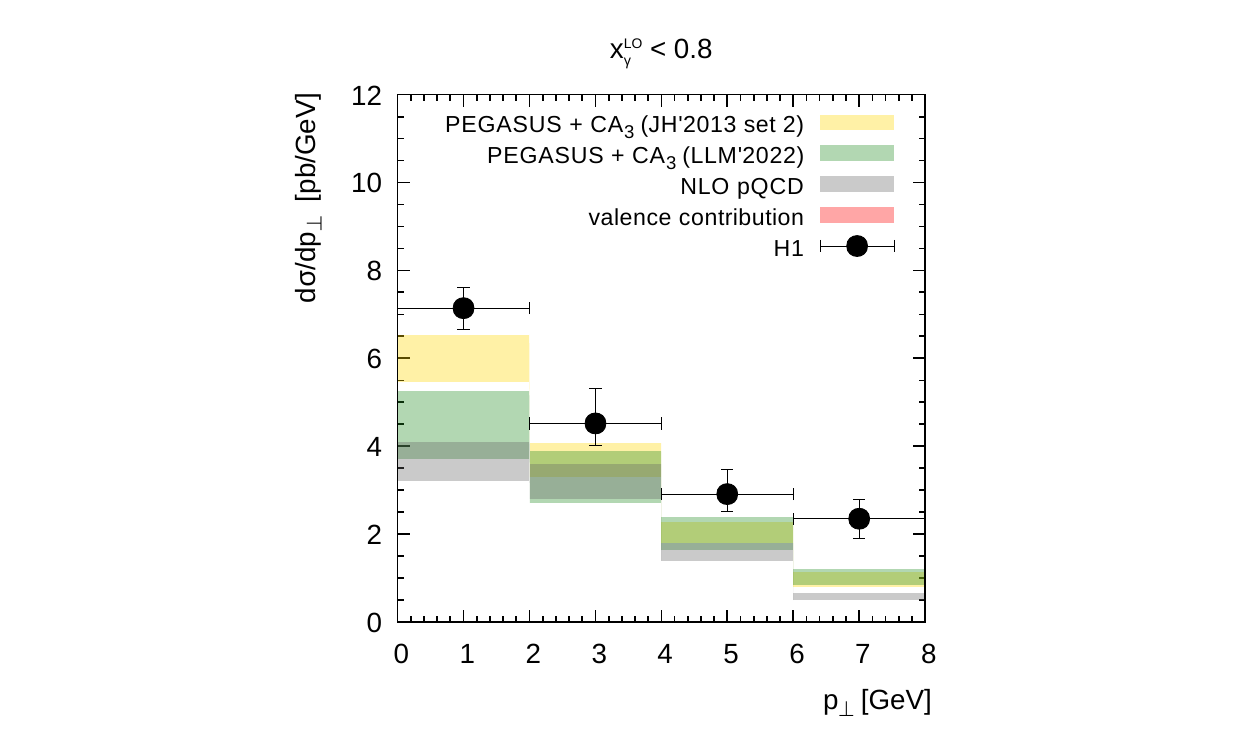}
\caption{The associated prompt photon and jet photoproduction cross
section as functions of photon and jet azimuthal angle difference $\Delta\phi$ and 
photon momentum perpendicular to the jet direction $p_\perp$ at $x_\gamma^{\rm LO} > 0.8$ (left panels) and $x_\gamma^{\rm LO} < 0.8$ (right panels).
The notations are the same as in Fig.~\ref{fig-H1}. The experimental data are from H1~\cite{PromptPhoton-H1}.}
\label{fig-H12} 
\end{center}
\end{figure}

Our results are compared with experimental data in Figs.~\ref{fig-H1} ---\ref{fig-H12}.
So, the transverse energy $E_T^\gamma$, $E_T^{\rm jet}$ and pseudorapidity $\eta^\gamma$, $\eta^{\rm jet}$ distributions 
of the produced photon and jet are shown in Figs.~\ref{fig-H1} and \ref{fig-ZEUS}.
We find that the calculations based on recently proposed LLM'2022 gluon density 
(represented by the green histograms)
are consistent with the H1 data in most bins, although some discrepancies are
present. In particular, these predictions tend to underestimate the H1 measurements
at low $E_T^{\rm jet}$ and rear photon pseudorapidity $\eta^\gamma$,
although coincide with the data within the theoretical and experimental uncertainties.
Note that theoretical uncertainties of our calculations
are estimated in a traditional way, varying the renormalization scale\footnote{In the CCFM-based approach, 
the factorization scale $\mu_F$ is related with the evolution variable and therefore should not be varied. See\cite{CCFM} for more details.} 
around its default value as $\mu_{R}\to\xi\mu_{R}$ with $\xi = 1/2$ or $2$. 
One can see that the obtained results reproduce behavior of the measured $\eta^{\rm jet}$ spectrum
(shifted towards positive pseudorapidities),
which could not be achieved in earlier calculations\cite{LMZ-photo,KNS-photo}.
This is a direct consequense of the applied method of proper determination of 
jet kinematics based on the TMD shower algorithm.
Similar conclusions were already done in~\cite{LMJ-PP, LM-Higgs} where production 
of prompt photons or Higgs bosons associated with hadronic jet(s) in $pp$ collisions at the LHC was studied.
Overall agreement of our predictions with more recent ZEUS data is a bit worse, but still rather 
reasonable (see Fig.~\ref{fig-ZEUS}).
Although there is some overestimation of the data at low $E_T^\gamma$, large $E_T^{\rm jet}$
and low separation in azimuthal angle between the produced photon and jet, $\Delta \phi$,
the shapes of $\eta^{\rm jet}$ and $\Delta \eta = \eta^\gamma-\eta^{\rm jet}$
spectra are reproduced well.
These observables are sensitive to the proper determination of jet kinematics.

As it has been mentioned above, to investigate the dependence of our results
on the TMD gluon density in a proton, we have repeated the calculations
using another set, JH'2013 set 2.
In contrast with LLM'2022, it leads to systematic 
overestimation of the HERA data, that
coincides with the observation of~\cite{LLM-2023}.
It was argued\cite{LLM-2022, LLM-2023} that
better agreement achieved with the LLM'2022
is an immediate consequence of using a physically motivated expression
for the corresponding starting distribution.
Thus, our calculations demonstrate that the HERA data on 
associated $\gamma + {\rm jet}$ production in DIS are 
sensitive to the TMD gluon densities
and could help to clearly distinguish the latter.
It could be important for experiments at future electron-proton or electron-ion colliders,
such as LHeC, FCC-eh, EiC and EiCC.
Additionally, we find that contribution from the valence quarks
is negligible in the considered kinematical region.

For comparison we also show results of conventional NLO pQCD calculations 
taken from\cite{PromptPhoton-H1, PromptPhoton-ZEUS1, PromptPhoton-ZEUS2}. 
These predictions are slightly below the LLM'2022 ones but rather close to them for most 
of observables within the uncertainties. 
The NLO pQCD calculations 
also tend to underestimate the H1 data.  However, they agree well with both the H1 and ZEUS
measurements at large $E_T^{\rm jet}$ and small $\Delta \phi < 90^\circ$.
The JH'2013 set 2 gluon density significantly overshoots the NLO pQCD results.

Other important variables are the 
longitudinal momenta fractions carried by the colliding partons. 
The momentum fractions of the initial photon and proton are introduced in the ZEUS analyses\cite{PromptPhoton-ZEUS1, PromptPhoton-ZEUS2} as the following:
\begin{equation}
x_\gamma^{\rm obs}=\frac{E^\gamma_T e^{-\eta^\gamma}+E_T^{\rm jet}e^{-\eta^{\rm jet}}}{2yE_e}, \quad x_p^{\rm obs}=\frac{E^\gamma_T e^{\eta^\gamma}+E_T^{\rm jet}e^{\eta^{\rm jet}}}{2E_p}.
\end{equation}
\noindent
At $x_\gamma^{\rm obs}>0.8$ the cross section is believed to be dominated by 
the "direct photon" contributions, whereas at lower $x_\gamma^{\rm obs}$ the "resolved photon" contributions play the main role.
The H1 Collaboration refers to $x_\gamma^{\rm LO}$ and $x_p^{\rm LO}$ estimators given by\cite{PromptPhoton-H1}:
\begin{equation}
x_\gamma^{\rm LO}=\frac{E_T^\gamma(e^{-\eta^\gamma}+e^{-\eta^{\rm jet}})}{2yE_e},\quad x_p^{\rm LO}=\frac{E_T^\gamma(e^{-\eta^\gamma}+e^{\eta^{\rm jet}})}{2E_p}.
\end{equation}
\noindent
The $x_\gamma^{\rm LO}$ and $x_p^{\rm LO}$ variables 
explicitly use only of the photon energy, which is better measured
than the jet energy. 
Our predictions for these observables are shown in Fig.~\ref{fig-x}.
We find that results obtained with
LLM'2022 gluon density agree well with the H1 data.
In contrast with earlier calculations\cite{LMZ-photo}, they show 
more smeared cross sections for the $x_\gamma^{\rm obs}$ and $x_\gamma^{\rm LO}$ spectra, 
which is in a better agreement with the data.
Again, this is  due to more accurate determination of the 
jet kinematics compared to the previous considerations.
However, there is still some overestimation of the ZEUS data at small $x_\gamma^{\rm obs}$ and large $x_p^{\rm obs}$.

Although we cannot distinguish between the "resolved photon" 
and "direct photon" contributions, 
the dedicated study of cross sections measured at low and high 
$x_\gamma^{\rm obs}$
can provide an additional test of theoretical calculations. 
In fact, in our simulations different interplay between the subprocesses (\ref{eq:vg2vqq}) and 
(\ref{eq:vg2vg}) can result then in different cross sections in these two regions 
and thus there might be a deviation from the data. 
To investigate it in more details,
we have performed the calculations at low $x_\gamma^{\rm obs}(x_\gamma^{\rm LO}) < 0.8$ and high $x_\gamma^{\rm obs}(x_\gamma^{\rm LO}) > 0.8$ values
and compared our results with the H1\cite{PromptPhoton-H1} and ZEUS\cite{PromptPhoton-ZEUS2} measurements, see Figs.~\ref{fig-ZEUS2-pt} --- \ref{fig-H12}.
One can see that transverse energy $E_T^\gamma$, $E_T^{\rm jet}$ 
and pseudorapidity $\eta^\gamma$, $\eta^{\rm jet}$ 
spectra predicted by the LLM'2022 gluon density
are in a reasonable agreement with the ZEUS data\cite{PromptPhoton-ZEUS2}
in the both these kinematical regions within the uncertainties,
except only last $E_T^{\rm jet}$ 
and forward $\eta^{\rm jet}$ bins at $x_\gamma^{\rm obs} < 0.8$.
However, there is some overestimation of the ZEUS data at large $x_p^{\rm obs} \geq 0.015$, 
that is clearly visible at $x_\gamma^{\rm obs} < 0.8$.
It comes from the events where the produced photon
and jet are close to each other, as one can see from $\Delta \phi$ distributions shown in Fig.~\ref{fig-ZEUS2-oth}.
Note that our predictions overshoot the ZEUS data on $\Delta \phi$ spectra 
at $\Delta \phi < 150^\circ$ and $x_\gamma^{\rm obs} > 0.8$, but agree well with the H1 measurements
which were performed in the similar kinematical region.
In contrast, at low $x_\gamma^{\rm obs} < 0.8$ the H1 data for $\Delta \phi$ distributions 
are clearly underestimated everywhere while corresponding ZEUS data are reasonably well described.
Therefore, at this point it could be some contradiction between the H1 and ZEUS measurements.
In any case, special studies of transverse correlations between the final state particles are known to be useful to 
investigate the production dynamics (see, for example,\cite{LZ-photo1, LZ-photo2, LMZ-photo, LZ-PP} and references therein).
So, the H1 Collaboration has investigated the distribution
on the component of the photon's momentum perpendicular to 
the jet direction in the transverse plane, 
$p_\perp = |{\mathbf p}_T^\gamma \times {\mathbf p}_T^{\rm jet}|/|{\mathbf p}_T^{\rm jet}| = E_T^\gamma \sin \Delta \phi$.
Similar to $\Delta \phi$ spectra, the distribution over $p_\perp$ 
is partucularly sensitive to the higher-order pQCD corrections\footnote{In the conventional (collinear) LO approximation, 
it must be simply a delta function since the produced photon and the jet are 
back-to-back in the transverse plane.}, which are taken into 
account in the form of CCFM-evolved TMD gluon densities in our calculations.
The predictions for $p_\perp$ spectra are confronted with the H1 data
in Fig.~\ref{fig-H12}. We find that $k_T$-factorization results 
agree well with the H1 data in the "direct photon" region,
although LLM'2022 gluon slightly underestimates the data at low $p_\perp$ and
JH'2013 set 2 tends to overshoots the latter.
The conventional NLO pQCD calculations do not reproduce the H1 data
both in the normalization and shape.
At the same time we find that none of the calculations is able to describe the overall 
normalization of the data at $x_\gamma^{\rm LO} < 0.8$,
although the shape of measured distribution is reproduced well by all of them.
Thus, further investigation of such an observable
could be important to discriminate between the different approaches.

Finally, we can conclude that the LLM'2022 predictions
reproduce well the latest HERA data for most of the observables
within the theoretical and experimental uncertainties. 
It is due to more accurate determination of jet kinematics in 
our consideration compared to earlier analyses performed within the TMD-based approaches.
These calculations are rather close to the NLO pQCD ones at $x_\gamma^{\rm obs} > 0.8$ 
and tend to lie above the latter at $x_\gamma^{\rm obs} < 0.8$.
The NLO pQCD provides a better description of the jet transverse energy
spectra, but unable to reproduce well the data on most of correlation variables (such as $p_\perp$ ones).
The JH'2013 set 2 gluon density overshoots the HERA data,
although at $x_\gamma^{\rm obs} < 0.8$ the difference between 
these predictions and LLM'2022 ones become smaller.
So, the cross sections of associated $\gamma + {\rm jet}$ production in DIS events
are sensitive to the TMD gluon density in a proton and can be used to constrain it.

\section{Conclusion} \indent

We have considered associated production of prompt photon and hadronic jets 
in photoproduction regime of deep inelastic electron-proton scattering at high energies.
The calculations were performed in the framework of $k_T$-factorization QCD approach
and mainly based on two off-shell photon-gluon fusion subprocesses, 
$\gamma + g^* \to \gamma + q + \bar q$ and $\gamma + g^* \to \gamma + g$,
implemented now into the Monte-Carlo generator \textsc{pegasus}.
First of them, being supplemented with effects of the 
final-state parton radiation, covers many other subprocesses, including resolved photon contributions.
An additional valence quark-induced subprocess $\gamma + q_{\rm val} \to \gamma + q$, which 
can be important at large $x$ region, has been 
taken into account in the conventional (collinear) QCD factorization.
In the numerical calculations we have tested two CCFM-evolved gluon distributions
in a proton, namely, JH'2013 set 2 and very recent LLM'2022 gluon, both
available in \textsc{pegasus} and \textsc{tmdlib} packages.
The LLM'2022 gluon density is based on simultaneous fit to the number of HERA and LHC processes
sensitive to the gluon content of the proton.
To reconstruct correctly the kinematics of the hadronic jets the TMD
parton shower implemented into the Monte-Carlo generator \textsc{cascade} has been applied for the first time for the DIS process.

We have achieved reasonably good agreement between our predictions obtained with the recent 
LLM'2022 gluon density in a proton and latest H1 and ZEUS experimental data, thus demonstrating again the importance 
of initial state TMD parton showers for jet determination in the TMD-based approaches.
The previously developed framework to calculate the jet associated 
processes is extended now to the electron-proton deep inelastic scattering.
It is important for forthcoming studies at future 
electron-proton and electron-ion colliders, 
such as LHeC, FCC-he, EiC and EiCC.

The next version of Monte-Carlo event generator \textsc{pegasus} (1.08), which is extended now to the considered DIS process, will be released shortly.

\section*{Acknowledgements}
We thank S.P.~Baranov and G.I. Lykasov for their important 
comments and remarks. 
We are also grateful to H.~Jung for help with implementation of \textsc{cascade} in DIS.
Framework setting, matrix elements adaptation and the analysis of the results were supported by the grant of the Foundation for the
Advancement of Theoretical Physics and Mathematics “BASIS” 20-1-3-11-1.
Work on implementation of parton showers and hadronization for photoproduction processes were supported by the 
Russian Science Foundation under grant~22-22-00119.


\newpage 

\end{document}